\documentclass[10pt]{article}
\usepackage{graphics,color}
\usepackage{bm}
\usepackage{graphicx}
\usepackage{amsmath}
\usepackage{amssymb}
\usepackage{subfigure}
\usepackage{authblk}
\usepackage[margin=0.68in,footskip=0.2in]{geometry}
\usepackage{tikz}
\usepackage{enumitem}
\usepackage{tensor}	
\usepackage{epstopdf}
\usepackage{hyperref}
\usepackage[normalem]{ulem}
\usepackage{soul}

\def\be{\begin{equation}}
\def\ee{\end{equation}}
\def\beq{\begin{equation}}
\def\eneq{\end{equation}}
\def\bea{\begin{eqnarray}}
\def\eea{\end{eqnarray}}

\newcommand{\bear}{\begin{eqnarray}}
\newcommand{\eear}{\end{eqnarray}}

\newlength{\tskip}
\newbox\pippobox

\def\9{\nabla}
\def\r{\rho}

\def\nn{\nonumber}

\def\6{\partial}

\def\f{\frac}

\def\0{(0)}

\def\>{\rightarrow}

\def\ba{\begin{eqnarray}}
\def\ea{\end{eqnarray}}
\def\l{\left}
\def\r{\right}
\def\f{\frac}

\def\hub{{\mathcal H}}

\title{{\bf Ho\v rava Gravity in the Effective Field Theory formalism: \\ from cosmology to observational constraints}}

\author[1]{Noemi Frusciante\thanks{E-mail:fruscian\textit{@}iap.fr}}
\author[2,3,4]{Marco Raveri\thanks{E-mail:mraveri\textit{@}sissa.it}}
\author[1]{Daniele Vernieri\thanks{E-mail:vernieri\textit{@}iap.fr, {\bf corresponding author}}}
\author[5]{Bin Hu\thanks{E-mail:hu\textit{@}lorentz.leidenuniv.nl}}
\author[5]{Alessandra Silvestri\thanks{E-mail:silvestri\textit{@}lorentz.leidenuniv.nl}}
\affil[1]{Sorbonne Universit$\acute{\text{e}}$s, UPMC Univ Paris 6 et CNRS, UMR 7095, Institut d'Astrophysique de Paris, GReCO, 98 bis bd Arago, 75014 Paris, France} 
\affil[2]{SISSA - International School for Advanced Studies, Via Bonomea 265, 34136, Trieste, Italy}
\affil[3]{ INFN, Sezione di Trieste, Via Valerio 2, I-34127 Trieste, Italy}
\affil[4]{ INAF-Osservatorio Astronomico di Trieste, Via G.B. Tiepolo 11, I-34131 Trieste, Italy }
\affil[5]{Institute Lorentz, Leiden University, PO Box 9506, Leiden 2300 RA, The Netherlands}

\begin{document}

\maketitle

\begin{abstract}
We consider Ho\v rava gravity within the framework of the effective field theory (EFT)
of dark energy and modified gravity. We  work out a complete mapping of the theory
into the EFT language for an action including all the operators which are relevant
for linear perturbations with up to sixth order spatial derivatives. We then employ
an updated version of the EFTCAMB/EFTCosmoMC package to study the cosmology  of the
low-energy limit of Ho\v rava gravity and place  constraints on its parameters using
several cosmological data sets.  In particular we use cosmic microwave background
(CMB) temperature-temperature and lensing power spectra by \emph{Planck} 2013, WMAP low-$\ell$
polarization spectra, WiggleZ galaxy power spectrum, local Hubble measurements,
Supernovae data  from SNLS, SDSS and HST and the baryon acoustic oscillations
measurements from BOSS, SDSS and 6dFGS. We get  improved upper bounds, with respect
to those from Big Bang Nucleosynthesis, on the deviation of the cosmological
gravitational constant from the local Newtonian one. At the level of the background
phenomenology, we find a relevant rescaling of the Hubble rate at all epoch, which
has a strong impact on the cosmological observables; at the level of perturbations,
we discuss in details all the relevant effects on the observables and find that in
general the quasi-static approximation is not safe to describe the evolution of
perturbations.
Overall we find that  the effects of the modifications induced by the low-energy
Ho\v rava gravity action are quite dramatic and current data place tight bounds on the theory
parameters.

\end{abstract}


\maketitle
\newpage

\tableofcontents

%
\section{Introduction}\label{Sec:Intro}

In their quest to find a  quantum theory of gravity that could describe physical phenomena at the Planck scale ($\sim 10^{19} \,\, \text{GeV}/\text{c}^2$),  relativists have recently started to explore Lorentz violating theories (LV) (see~\cite{Liberati:2013xla} and references therein). Indeed, even though Lorentz invariance (LI) is considered a cornerstone of our knowledge of reality, the challenge presented by physics at Planck energy is forcing us to question also our firmest assumptions. In the cosmological context,  LV theories represent interesting candidates for cosmic acceleration, since in their low-energy limit they generally predict a dynamical scalar degree of freedom (DoF) which could  provide a source for the late time acceleration, in alternative to the cosmological constant. While the standard model of cosmology, based on the laws of General Relativity (GR), is to date a  very good fit to available data, some outstanding theoretical problems related to the cosmological constant have indeed led people to explore alternative theories. To this extent, a wide range of models have been proposed, which either introduce a dynamical dark energy (DE) or modify the laws of gravity on large scales (MG) in order to achieve self accelerating solutions in the presence of negligible matter. All these alternatives generally result in the emergence of new scalar dynamical DoF (see~\cite{Silvestri:2009hh,Clifton:2011jh,Joyce:2014kja} for a comprehensive review), as it is the case with LV theories. 

Interestingly, LV theories typically break LI at all scales, and are therefore constrainable with many different measurements and data sets over a vast range of energies.
Constraints and measurements on the parameters of a general realistic effective field theory for Lorentz violation~\cite{Kostelecky:2008ts}, usually referred to as the Standard Model Extension~\cite{Colladay:1996iz,Colladay:1998fq},  support LI with an exquisite accuracy. Furthermore, LI has been tested to high accuracy on solar system scales, and stringent bounds have been placed on the Post Newtonian parameters (PPN), in particular on those corresponding to the preferred frame effects, since such effects are typical of LV theories~\cite{Will:2014kxa}. Phenomena on astrophysical scales, and in particular tests of gravity in the strong regime,  such as those of binary pulsars~\cite{Yagi:2013qpa,Yagi:2013ava}, provide further bounds on LV~\cite{Will:2014kxa}. On the contrary, the exploration of cosmological bounds on LV theories is still in its infancy~\cite{Carroll:2004ai,Zuntz:2008zz,Audren:2013dwa,Blas:2012vn,Audren:2014hza}.

In the present work, we focus on the class of LV theories known as Ho\v rava gravity~\cite{Horava:2008ih,Horava:2009uw} which modifies the gravitational action by adding higher order spatial derivatives without adding higher order time derivatives, thus modifying the graviton propagator and achieving a power-counting renormalizability. This is  possible if one considers that space and time scale differently. Such a prescription is implemented through a breaking of full diffeomorphism invariance, which leads to LV at all scales. The resulting theory propagates a new dynamical scalar DoF, i.e. the spin-0 graviton. As a  candidate for quantum gravity, Ho\v rava theory is expected to be renormalizable and also unitary. Nevertheless, at the moment there is no evidence for renormalizability beyond the power-counting arguments. 

Ho\v rava gravity shows a rich phenomenology on cosmological scales, e.g. the higher curvature terms in the action lead to a matter bouncing cosmology~\cite{Calcagni:2009ar,Kiritsis:2009sh}; it also shows different  mechanisms by which it is possible to explain the nearly scale invariant spectrum of cosmological perturbations without introducing an inflationary phase~\cite{Brandenberger:2009yt,Mukohyama:2009gg,Cai:2009dx,Chen:2009jr,Cai:2010hi}, finally, cosmological perturbations at late time have been investigated in refs.~\cite{Carroll:2004ai,Gao:2009ht,Wang:2009yz,Kobayashi:2009hh,Kobayashi:2010eh,Dutta:2009jn,Dutta:2010jh}.

In this paper, we perform a thorough analysis of the cosmology in Ho\v rava gravity by mapping the theory into the framework of Effective Field Theory (EFT) of cosmic acceleration  developed in refs.~\cite{Gubitosi:2012hu,Bloomfield:2012ff,Gleyzes:2013ooa,Bloomfield:2013efa,Piazza:2013coa,Frusciante:2013zop,Gleyzes:2014rba},  on the line of the EFT of inflation and quintessence~\cite{Cheung:2007st,Weinberg:2008hq,Creminelli:2008wc}. The basic idea of this framework  is to construct an effective action with  all the operators which are of relevance to study linear cosmological perturbations  around a Friedmann-Lema$\hat{\text{i}}$tre-Robertson-Walker (FLRW) background  and are invariant under time-dependent spatial diffeomorphisms. Indeed an expanding FLRW background breaks time-dependent diffeomorphism, allowing all  these operators to enter the action and, furthermore, to be multiplied by a free function of time~\cite{Cheung:2007st,Piazza:2013coa}. The resulting action encompasses most models of single scalar field DE and MG which have a well defined  Jordan frame.  In refs.~\cite{Hu:2013twa,Raveri:2014cka,Hu:2014sea,Hu:2014oga}, the EFT framework has been implemented in the public Einstein-Boltzmann solver CAMB~\cite{CAMB,Lewis:1999bs}, and the associated Monte Carlo Markov Chain code CosmoMC~\cite{Lewis:2002ah}. The resulting patches, dubbed EFTCAMB/EFTCosmoMC, are now publicly available at \url{http://wwwhome.lorentz.leidenuniv.nl/~hu/codes/} and represent a powerful package which allows to explore cosmological constraints both in a model independent and model specific way~\cite{Hu:2013twa}. The original action considered in ref.~\cite{Bloomfield:2012ff}, and implemented in the public version of EFTCAMB, contains all Horndeski and some of the extensions like GLPV~\cite{Gleyzes:2014rba,Kase:2014cwa,Gleyzes:2014dya}, but  does not have all the  operators necessary to study Ho\v rava gravity. The inclusion of Ho\v rava gravity in the context of EFT of DE/MG has been recently considered and investigated in refs.~\cite{Kase:2014cwa,Gao:2014soa}.  In this paper, we consider the most general action for Ho\v rava gravity with all the operators with up to sixth order spatial derivatives, which is  the minimal prescription to achieve power counting renormalizability. We focus on the part of this action that contributes to  linear order in perturbations~\cite{Blas:2009qj}. For this action we work  out a complete mapping to the EFT framework deriving also the generalization of the original EFT action used in refs.~\cite{Hu:2013twa,Gleyzes:2014rba}.  When we compare the predictions of the theory to the observations, we consider only the low-energy operators of Ho\v rava gravity, since those are the relevant ones to describe the large scale cosmology associated to the observables that we employ. We work out the contribution of these operators to the equations of motion for linear scalar and tensor perturbations, implementing them in an updated version of EFTCAMB that will be publicly released in the near future. 

The structure of the paper is the following. In section~\ref{theory}, we set up the theoretical background of the paper. In particular, in section~\ref{horava}, we  introduce  Ho\v rava gravity  and its main features, while  in section~\ref{eft}, we summarize the EFT framework and its implementation in EFTCAMB/EFTCosmoMC. In section~\ref{mapping} we work out the mapping of  Ho\v rava gravity in the EFT language focusing on  the low-energy part of the action and leaving the mapping of the high-energy part of the action to appendix~\ref{L4L6}. Finally, in section~\ref{Stability},  we discuss the requirements that EFTCAMB enforces on the scalar and tensor DoFs to prevent instabilities in the theory. In section~\ref{Sec:HoravaCosmology}, we study the cosmology of Ho\v rava gravity, discussing in detail how the model is implemented in EFTCAMB and what are the general effects of the modifications on the background and the perturbations . Finally, in section~\ref{Sec:CosmologicalConstraints} we explore observational constraints from several combinations of cosmological data sets. To this extent we consider two cases: the  low-energy Ho\v rava gravity action which is characterized by three constant parameters; a subcase of the latter, that evades PPN constraints and is characterized by two parameters. We draw our conclusions in section~\ref{conclusion}, discussing the main results. 

\section{Theory}\label{theory}

In this section we set up the theoretical basis for our analysis. In section~\ref{horava}, we introduce the main aspects of Ho\v rava gravity, which is the theory we want to  investigate and constrain by using the EFT approach. In section~\ref{eft}, we review the EFT framework, discussing its implementation in EFTCAMB, which is the Einstein-Boltzmann solver we use to perform a thorough investigation of the cosmology of the theory. In section~\ref{mapping}, we work out the mapping of the low-energy Ho\v rava gravity action in terms of the EFT functions. The mapping of the high-energy part of the action is discussed in appendix~\ref{L4L6}.  Finally, in section~\ref{Stability} we present the full set of equations evolved by EFTCAMB and the conditions that we impose on the tensor and scalar DoFs to ensure that the theory we are considering is viable.

\subsection{Ho\v rava Gravity} \label{horava}

Ho\v rava gravity has been recently proposed as a candidate for an ultraviolet completion of GR~\cite{Horava:2008ih,Horava:2009uw}. The basic idea is to modify the graviton propagator by adding to the action higher-order spatial derivatives without adding higher-order time derivatives, in order to avoid the presence of Ostrogradski instabilities~\cite{Ostrogradski}. The theory is constructed in such a way to be compatible with a different scaling of space and time, i.e. 
\beq
\left[dt\right] = \left[k\right]^{-z}, \qquad \left[dx\right] = \left[k\right]^{-1},
\eneq
where $z$ is a positive integer and $k$ is the momentum.
In order to accommodate such a different scaling between space and time, the action of Ho\v rava gravity cannot still be invariant under the full set of diffeomorphisms as in GR, but it can be invariant under the more restricted foliation-preserving diffeomorphisms
\beq
t \rightarrow \tilde{t}\left(t\right), \qquad x^i \rightarrow \tilde{x}^i\left(t,x^i\right). 
\eneq
Therefore, within this approach, space and time are naturally treated on different footing leading to Lorentz violations at all scales. The emergence of LV is reflected in modified  dispersion relations for the propagating DoFs. From a practical point of view, the different behavior of space and time is achieved by picking a preferred foliation of spacetime, geometrically described within the Arnowitt-Deser-Misner (ADM) formalism.

It has been shown that the theory is power-counting renormalizable if and only if $z\geq d$, where $d$ indicates the number of spatial dimensions, which means that the action has to contain operators with at least $2d$ spatial derivatives~\cite{Visser:2009fg,Visser:2009ys}. Hence, in a four-dimensional spacetime, $d=3$,  power-counting renormalizability arguments request at least sixth-order spatial derivatives in the action. 

Considering the above arguments, the action of Ho\v rava gravity can be written as follows~\cite{Blas:2009qj}
\begin{eqnarray} \label{horavaaction}
\mathcal{S}_{H}=\f{1}{16\pi G_H}\int{}d^4x\sqrt{-g}\left(K_{ij}K^{ij}-\lambda K^2 -2 \xi\bar{\Lambda} +\xi \mathcal{R}+\eta a_i a^i +L_4+L_6\right)+S_m[g_{\mu\nu},\chi_i],
\end{eqnarray}
where $g$ is the determinant of the metric $g_{\mu\nu}$, $\mathcal{R}$ is the Ricci scalar of the three-dimensional space-like hypersurfaces, $K_{ij}$ is the extrinsic curvature, and $K$ is its trace. $\left\{\lambda,\xi,\eta\right\}$ are dimensionless running coupling constants, $\bar{\Lambda}$ is the ``bare'' cosmological constant, $a_i=\partial_i \mbox{ln} N$ where as usual $N$ is the lapse function of the ADM metric. $L_4$ and $L_6$ denote the Lagrangians associated to the higher-order operators, that contain, respectively, fourth and sixth-order spatial derivatives (see appendix~\ref{L4L6} for the explicit expressions of their parts that  contribute to  linear order perturbations). These Lagrangians constitute the high-energy (HE) part of the action~(\ref{horavaaction}),  while the operators preceding them represent the low-energy (LE) limit of the theory and are the ones of relevance on large scale. $S_m$ is the matter action for all matter fields, $\chi_i$. Finally, $G_H$ is  the coupling constant which can be expressed as 
\be
G_H=\xi G
\ee   
where $G$ is the ``bare'' gravitational constant. As demonstrated in ref.~\cite{Blas:2009qj}, the solution of the static point-like mass in the Newtonian limit gives the relationship between the ``bare'' gravitational constant ($G$) and the Newtonian one ($G_N$), i.e.
\be
G=G_N\l(1-\f{\eta}{2\xi}\r).
\ee
Then, the coupling in front of the action reads 
\be
\f{1}{16\pi G_H}=\f{m_0^2}{(2\xi-\eta)},
\ee
where  $m_0^2=1/8\pi G_N$ is the Planck mass defined  locally.

Notice that the action of GR is recovered when $\lambda=1$, $\xi=1$ and $\eta=0$, and the higher order operators in $L_4$ and $L_6$ are not considered. 

The symmetry of the theory allows for a very large number of operators $\sim\mathcal{O}(10^2)$ in $L_4$ and $L_6$.  In order to limit the huge proliferation of couplings in the full theory, in the first proposal Ho\v rava imposed some restrictions, i.e. projectability and detailed balance (for the details see refs.~\cite{Sotiriou:2009gy,Sotiriou:2009bx,Sotiriou:2010wn,Visser:2011mf,Vernieri:2011aa,Vernieri:2012ms,Vernieri:2015uma}). In the following we will not impose any of these limitations to the action~(\ref{horavaaction}) and we will consider for $L_4$ and $L_6$ all the operators which contribute to the dynamics of linear perturbations~\cite{Blas:2009qj}. 

\subsection{Effective Field Theory Framework}\label{eft}
In the effective field theory approach to DE/MG~\cite{Gubitosi:2012hu,Bloomfield:2012ff}, an action is built in the Jordan frame and unitary gauge by considering  the operators which are invariant under time-dependent spatial diffeomorphisms. The additional scalar DoF representing DE/MG is eaten by the metric via a foliation of space-time into space-like hypersurfaces which correspond to a uniform scalar field. At quadratic order, which is sufficient to study the dynamics of linear perturbations, the action reads
\begin{align}
\mathcal{S}_{EFT} = \int d^4x &\sqrt{-g}  \bigg\{ \frac{m_0^2}{2} \left[1+\Omega(\tau)\right] R + \Lambda(\tau) - c(\tau)\,a^2\delta g^{00}  
  + \frac{M_2^4 (\tau)}{2} \left( a^2\delta g^{00} \right)^2
 - \frac{\bar{M}_1^3 (\tau)}{2} \, a^2\delta g^{00}\,\delta \tensor{K}{^\mu_\mu}  
	\nonumber \\
   & - \frac{\bar{M}_2^2 (\tau)}{2} \left( \delta \tensor{K}{^\mu_\mu}\right)^2
    - \frac{\bar{M}_3^2 (\tau)}{2} \,\delta \tensor{K}{^\mu_\nu}\,\delta \tensor{K}{^\nu_\mu}
      + m_2^2(\tau)\left(g^{\mu\nu}+n^{\mu} n^{\nu}\right)\partial_{\mu}(a^2g^{00})\partial_{\nu}(a^2g^{00}) \nonumber \\
	  & +\frac{\hat{M}^2(\tau)}{2} \, a^2 \delta g^{00}\,\delta \mathcal{R}+	\ldots \bigg\} + S_{m} [g_{\mu \nu}, \chi_m ],\label{actioneft}
\end{align}
where $R$ is the four-dimensional Ricci scalar, $\delta g^{00}$, $\delta \tensor{K}{^\mu_\nu}$, $\delta \tensor{K}{^\mu_\mu}$ and  $\delta \mathcal{R}$ are respectively the perturbations of the upper time-time component of the metric, the extrinsic curvature and its trace and the  three dimensional spatial Ricci scalar. Finally,  $S_m$ is the matter action. Since the choice of the unitary gauge breaks time diffeomorphism invariance, each operator in the action can be multiplied by a time-dependent coefficient; in our convention, $\{\Omega,\Lambda,c, M_2^4,\bar{M}_1^3,\bar{M}_2^2,\bar{M}_2^2,\bar{M}_3^2,m_2^2,\hat{M}^2\}$ are unknown functions of the conformal time, $\tau$, and we will refer to them as EFT functions. In particular, $\{\Omega,c,\Lambda\}$ are the only functions contributing  both to the dynamics of the background  and  of the perturbations, while the others  play a role only at level of  perturbations.   Let us notice that the above action includes explicitly all the operators that in ref.~\cite{Bloomfield:2012ff} have been considered to be relevant for linear cosmological perturbations since they can be easily related to some well known DE/MG models such as f(R), quintessence, Horndeski, or because they have been already studied in the EFT of inflation~\cite{Weinberg:2008hq,Creminelli:2008wc,Cheung:2007st}.  For such operators the corresponding field equations have been worked out~\cite{Bloomfield:2012ff,Hu:2013twa}.  However, additional second order operators can also be considered, such as $(\delta \mathcal{R})^2$, $\delta \tensor{\mathcal{R}}{^i_j}\delta \tensor{\mathcal{R}}{^j_i}$ as well as operators with higher-order spatial derivatives acting on them,~\cite{Gubitosi:2012hu,Bloomfield:2012ff,Gleyzes:2013ooa,Kase:2014cwa}. In particular, as we will show in appendix~\ref{L4L6}, additional operators are needed to describe Ho\v rava gravity in the EFT framework (see also~\cite{Kase:2014cwa}).

As mentioned in the Introduction, action~(\ref{actioneft}) allows to describe in a unified language all single scalar field  dark energy and modified gravity models which have a  well defined  Jordan frame.  In unitary gauge the  extra scalar DoF is hidden inside the metric perturbations, however in order to study the dynamics of linear perturbations and  investigate the stability of a given model, it is convenient to make it explicit by means of  the St$\ddot{\text{u}}$kelberg technique i.e. performing an infinitesimal coordinate transformation such that $\tau\rightarrow \tau+\pi$, where the new field $\pi$ is the St$\ddot{\text{u}}$kelberg field which describes the extra propagating DoF. Correspondingly,  all the functions of time in action~(\ref{actioneft}) are expanded in Taylor-series and the operators transform accordingly to the tensor transformation laws~\cite{Gubitosi:2012hu,Bloomfield:2012ff}. Varying the action with respect to the $\pi$-field one obtains a dynamical perturbative equation for the extra DoF which allows to control directly the stability of the theory, as discussed at length in ref.~\cite{Hu:2013twa}.

In refs.~\cite{Hu:2013twa,Raveri:2014cka} the effective field theory framework has been implemented into CAMB/CosmoMC~\cite{CAMB,Lewis:1999bs,Lewis:2002ah} creating the EFTCAMB/EFTCosmoMC patches which are publicly available at \url{http://wwwhome.lorentz.leidenuniv.nl/~hu/codes/} (see ref.~\cite{Hu:2014oga} for technical details).  EFTCAMB evolves the full   equations for linear perturbations without relying on any quasi-static (QS) approximation. In addition to the standard matter components (i.e. dark matter, radiation and massless neutrinos), massive neutrinos have also been included~\cite{Hu:2014sea}.  EFTCAMB allows to study perturbations in a model independent way (usually referred to as \emph{pure} EFT mode), investigating the cosmological implications of the different operators in action~(\ref{actioneft}). It can also be used to study  the exact dynamics for specific models, after the mapping of the given model into the EFT language has been worked out (usually referred to as \emph{mapping} mode). In the latter case one can treat the background  via a designer approach, i.e. fixing the expansion history and reconstructing the specific model in terms of EFT functions; or one can  solve the full background equations of the chosen theory. We refer to the latter as the \emph{full mapping} case. Furthermore, the code has a powerful built-in module that investigates whether a chosen model is  viable,  through a set of  general conditions of mathematical and physical stability. In particular,  the physical requirements include the avoidance of ghost and gradient instabilities for both the scalar and the tensor DoFs. The stability requirements are translated into \emph{viability  priors} on the parameter space when using EFTCosmoMC to interface EFTCAMB with cosmological data, and they can sometimes dominate over the constraining power of data~\cite{Raveri:2014cka}.
In this paper we will  study the case of Ho\v rava gravity, first describing how it can be cast into EFTCAMB via a full mapping, then exploring the effects of the stability conditions on its parameter space and finally deriving constraints from different combinations of cosmological data sets.

\subsection{Mapping Ho\v rava Gravity into the EFT approach}\label{mapping}

In this section we will work out explicitly the mapping of the low-energy (LE) part of  action~(\ref{horavaaction}) into the EFT formalism described in the previous section. This is the part of the action for which we will explore cosmological constraints. We show the mapping for the high-energy (HE) part ($L_4$ and $L_6$) in the appendix~\ref{L4L6}.

We use the following conventions:  (-,+,+,+) for the signature of the  metric $g_{\mu\nu}$; the background is considered FLRW with $\kappa=0$;  dots are derivatives w.r.t. conformal time, $\tau$ and $\hub\equiv \dot{a}/a$ is the Hubble rate; we will use the superscript $^{(0)}$ for the background quantities; finally  we define a time-like unit vector, $n_{\mu}$ as  
\begin{align}
n_\mu = \frac{\partial_\mu t}{\sqrt{-g^{\alpha\beta}\partial_\alpha t \partial_\beta t}}, \hspace{1cm} \mbox{with} \qquad n_\mu n^\mu =-1,
\end{align}
which corresponds to the convention that we use for the normal vector to the uniform-field hypersurfaces in the EFT construction of the action~(\ref{actioneft})~\cite{Bloomfield:2012ff,Hu:2013twa}. In conformal time and at second order in perturbations, one has

\ba
&&n_{\mu}=\delta_\mu^0\l(1+\f{1}{2}a^2\delta g^{00}+\f{3}{8}(a^2\delta g^{00})^2\r), \\
&&n^{\mu}=g^{0\mu}\l(1+\f{1}{2}a^2\delta g^{00}+\f{3}{8}(a^2\delta g^{00})^2\r).\label{upn}
\ea
In the following, these relations will be often employed.
 
Let us first recall the  low-energy action, which can be  rewritten as:
\begin{eqnarray}\label{LEhorava}
\mathcal{S}_{H,LE}&=& \f{m_0^2}{(2\xi-\eta)}\int{} d^4x \sqrt{-g} \left(K_{ij} K^{ij} - \lambda K^2 + \xi \mathcal{R} -2 \xi\bar{\Lambda}+ \eta a_i a^i \right) \nn \\
&=&  \f{m_0^2}{(2\xi-\eta)}\int{} d^4x \sqrt{-g}\l(\xi R +(1-\xi) K^{ij}K_{ij}+ (\xi-\lambda)K^2
 -2 \xi\bar{\Lambda}+ \eta a_i a^i \r)+ \mbox{boundary terms} \,,
\end{eqnarray}
where the second line has been obtained by using the Gauss-Codazzi relation~\cite{Gourgoulhon:2007ue}.\\
In the following, we show how to rewrite every single term of the above action in the EFT formalism described by the action~(\ref{actioneft}), providing the mapping of the Ho\v rava gravity parameters into the EFT functions.
 \begin{itemize}
 \item $\underline{\f{m_0^2\xi}{(2\xi-\eta)}\l(R-2 \bar{\Lambda}\r)}$\\
 Comparing the above expression with the EFT action~(\ref{actioneft}), it is straightforward to deduce that  these two  terms contribute to the following EFT functions
\be
(1+\Omega)=\f{2\xi}{(2\xi-\eta)}, \qquad \Lambda= -2\f{m_0^2\xi}{(2\xi-\eta)}\bar{\Lambda}.
\ee
\item \underline{$\f{m_0^2}{(2\xi-\eta)}(\xi-\lambda) K^2$} \\
 In order to identify the relation between the EFT functions and the Ho\v rava gravity parameters we have to expand $K^2$ up to second order in perturbations as
\be
K^2=2K^{(0)}K+(\delta K)^2-K^{(0)2},
\ee
by using $K=K^{(0)}+\delta K$.
Comparing the above relation with the action~(\ref{actioneft}), it is straightforward to see that the last term gives contribution  to $\Lambda(\tau)$ and the second one to $\bar{M}_2^2 (\tau)$. The first term can be computed as follows~\cite{Gubitosi:2012hu}
\bea
\int{}d^4x\sqrt{-g}2K^{(0)}K 
&=&2\int{} d^4x\sqrt{-g}K^{(0)} \left(\nabla_{\mu} n^{\mu}\right)
=-2\int{}d^4x\sqrt{-g}\nabla_{\mu}K^{(0)}n^{\mu} \nn \\
&=&2 \int{}d^4x\sqrt{-g} \f{\dot{K}^{(0)}}{a}\left[1-\f{1}{2}(a^2\delta g^{00})
-\f{1}{8}(a^2 \delta g^{00})^2\right] \label{Kterm},
\eea
where we have integrated by parts the second line and we have used eq.~(\ref{upn}). The last line will give respectively its contribution to $\Lambda(\tau)$, $c(\tau)$ and $M_2^4(\tau)$. Then summarizing, the corresponding contributions to the  EFT functions from the $K^2$ term are
\bea
&&\Lambda(\tau)=-\f{m_0^2(\xi-\lambda)}{(2\xi-\eta)}\l(K^{(0)2}-2 \f{\dot{K}^{(0)}}{a}\r), \qquad c(\tau)=\f{m_0^2(\xi-\lambda)}{(2\xi-\eta)}\f{\dot{K}^{(0)}}{a}, \nn\\
&&M_2^4(\tau)=-\f{m_0^2(\xi-\lambda)}{2(2\xi-\eta)}\f{\dot{K}^{(0)}}{a}, \qquad \bar{M}_2^2 (\tau)=-\f{2 m_0^2}{(2\xi-\eta)}(\xi-\lambda).
\eea
\item $\underline{\f{m_0^2(1-\xi)}{(2\xi-\eta)}K_{ij}K^{ij}}$ \\
As before, we can   expand up to second order in perturbations the above operator and it can be written as
\be
K_{ij}K^{ij}= 2K_{ij}^{(0)}\delta K^{ij}+K^{ij(0)}K_{ij}^{(0)}+\delta K_{ij}\delta K^{ij},
\ee
where we have  used the spatial metric to raise the indices and the extrinsic curvature has been decomposed  into its background and first order perturbation parts, i.e. $K_{ij}=K_{ij}^{(0)}+\delta K_{ij}$.  Moreover, the first term can be written as
\be
2K_{ij}^{(0)}\delta K^{ij}=-2\f{\hub}{a}\delta K =-2\f{\hub}{a^2} \l(a K+3\hub\r),
\ee
where the term proportional to $K$ can be treated as in eq.~(\ref{Kterm}). Finally, in terms of the EFT functions this operator can be written as
\ba
&&\Lambda(\tau)=-\f{m_0^2(1-\xi)}{(2\xi-\eta)}\l[K^{ij(0)}K_{ij}^{(0)}+\f{2}{a^2}\l(\dot{\hub}-\hub^2\r)\r]\,, \qquad c(\tau)=-\f{m_0^2(1-\xi)}{(2\xi-\eta) a^2}(\dot{\hub}-\hub^2)\,\nn\\
&&M^4_2(\tau)=\f{m_0^2(1-\xi)}{2a^2(2\xi-\eta)}(\dot{\hub}-\hub^2)\,, \qquad \bar{M}_3^2= -2\f{m_0^2(1-\xi)}{(2\xi-\eta)}.
\ea
\item $\underline{\f{m_0^2}{(2\xi-\eta)}\eta \,a_i a^i}$ \\
Let us first write explicitly $a_i$ in terms of perturbations up to second order 
\be\label{aiexpansion}
a_i=\f{\partial_i N}{N}=-\f{1}{2}\f{\partial_i (a^2 g^{00})}{a^2g^{00}}= \f{1}{2}\partial_i \delta (a^2 g^{00})+ \mathcal{O}(2),
\ee 
where in the last equality we have used $a^2g^{00}=-1+a^2\delta g^{00}$ and then we have expanded in Taylor series. Then we get
\be
\f{m_0^2}{(2\xi-\eta)}\eta a_ia^i=\f{m_0^2}{4(2\xi-\eta)}\eta \f{\tilde{g}^{ij}}{a^2}\partial_i (a^2 \delta g^{00})\partial_j (a^2 \delta g^{00})\,,
\ee
where $\tilde{g}^{ij}$ is the background value of the spatial metric. In the EFT language the above expression corresponds to 
\be
m^2_2=\f{m_0^2 \eta}{4(2\xi-\eta)}.
\ee
\end{itemize}
Summarizing, we can map the low-energy action~(\ref{LEhorava}) of Ho\v rava gravity in the EFT language at the basis of EFTCAMB as follows:
\begin{align}\label{Horava_mapping}
&(1+\Omega)=\f{2 \xi}{(2\xi-\eta)}, \nn \\
&c(\tau)= -\f{m_0^2}{a^2(2\xi-\eta)}(1+2\xi-3\lambda)\l(\dot{\hub}-\hub^2\r),\nn \\
&\Lambda(\tau)=\f{2m_0^2}{(2\xi-\eta)}\l[-\xi\bar{\Lambda}-(1-3\lambda+2\xi)\l(\f{\hub^2}{2a^2}+\f{\dot{\hub}}{a^2}\r)\r], \nn\\
&\bar{M}_3^2= -\f{2m_0^2}{(2\xi-\eta)}(1-\xi),\nn \\
&\bar{M}_2^2 =-2\f{m_0^2}{(2\xi-\eta)}(\xi-\lambda), \nn \\
&m^2_2=\f{m_0^2 \eta}{4(2\xi-\eta)},\nn \\
&M_2^4(\tau)=\f{m_0^2}{2a^2(2\xi-\eta)}(1+2\xi-3\lambda)\l(\dot{\hub}-\hub^2\r) \,,\nonumber\\
&\bar{M}_1^3=\hat{M}^2=0,
\end{align}
where we have explicitly written the value of the extrinsic curvature and its trace on a flat FLRW background \footnote{For the low-energy action it is possible to obtain part of the mapping by following the method in ref.~\cite{Gleyzes:2013ooa}. However, one has to consider that our formalism and notation differ from the one in ref.~\cite{Gleyzes:2013ooa} because we are using conformal time, a different signature for the normal unit vector, a different notation for the EFT functions and  one more operator is included in our low-energy action: $a_\mu a^\mu$.}. 
The mapping of the high-energy part of the action can be found in appendix~\ref{L4L6}. 

\subsection{Degrees of freedom: dynamics and stability}\label{Stability}
After the full diffeomorphism invariance is restored by means of the St$\ddot{\text{u}}$ckelberg mechanism,  at the level of perturbations we have a dynamical equation for the scalar DoF represented by the St$\ddot{\text{u}}$ckelberg field $\pi$.  In the case of the low-energy limit of  Ho\v rava gravity that we are considering, this equation reads
\be
\label{eq:pi1}
\eta\ddot{\pi}+2\eta\hub\dot{\pi}+\l[(3\lambda-2\xi-1)(\mathcal H^2-\dot{\mathcal H})+\eta (\mathcal H^2+\dot{\mathcal H})\r]\pi+k^2\xi(\lambda-1)\pi+\xi(\lambda-1) k\mathcal Z+\f{(\xi-1)(2\xi-\eta)}{2k}\left[\frac{a^2(\rho_{i}+p_{i})}{m_0^2}v_{i}\right]=0,
\ee
where $\mathcal{Z}$ is the standard CAMB variable~\cite{CAMB,Lewis:1999bs,Lewis:2002ah} $\rho_i$, $p_i$ are the background density and pressure of matter components, and $v_i$ is the velocity perturbation of matter components.  The above equation is coupled with the following perturbative field equations:
\begin{itemize}
\item time-time  (t) field equation 
\be \label{ttcomponent}
2  \hub \l[k^2 \pi (\eta -3 \lambda +2 \xi +1)+(1-3 \lambda ) k\mathcal{Z}\r]+2 k^2   \left(2 \xi  \bar{\eta} +\eta  \dot{\pi}\right)+a^2\f{ 2 \xi-\eta }{m_0^2} \delta \rho_m=0\,, \\
\ee
\item  space-space  (s) field equation
\be\label{tracelesscomponent}
-4  \hub \l[k^2 (3 \lambda -2 \xi -1) \pi+(3 \lambda -1) k \mathcal{Z}\r]+(1-3 \lambda ) \ddot{h}+4 k^2 \xi  \bar{\eta}+2 k^2 (-3 \lambda +2 \xi +1) \dot{\pi}+3 a^2 \f{(\eta -2 \xi )}{m_0^2} \delta P_m=0\,,
\ee
\end{itemize}
where $h,\bar{\eta}$ are the usual scalar perturbations of the metric in synchronous gauge (notice that we have added a bar to the standard metric perturbation in order to do not confuse it with the Ho\v rava gravity parameter, $\eta$).  EFTCAMB  evolves the above set of coupled differential equations  along with the usual matter perturbation equations and the initial conditions are set following ref.~\cite{Hu:2013twa}.  Let us notice that by using the mapping~(\ref{Horava_mapping}) worked out in the previous section, it is straightforward to deduce the above equations following the general prescription in ref.~\cite{Hu:2013twa}. 

We shall now  determine  the dispersion relation of the scalar DoF,  computing the determinant of the matrix of the coupled system eqs.~(\ref{eq:pi1})-~(\ref{tracelesscomponent}). Since the number counting of dynamical DoFs will not be changed by neglecting the couplings with standard matter 
species, for simplicity, for the purpose of this calculation we neglect them. After taking the Fourier transform $\partial_{\tau}\rightarrow -i\omega$, we can rewrite the system~(\ref{eq:pi1})-~(\ref{tracelesscomponent}) in the following matrix form: 
\be
\left(
\begin{array}{ccc}
 \gamma _{\pi \pi } & \gamma _{\text{$\pi $h}} & \gamma _{\pi \bar\eta } \\
 \gamma _{\text{s$\pi $}} & \gamma _{\text{sh}} & \gamma _{\text{s$\bar\eta $}}\\
 \gamma _{\text{t$\pi $}} & \gamma _{\text{th}} & \gamma _{\text{t$\bar\eta $}} 
\end{array}
\right)\left( \begin{array}{cc} &\pi \\ &h \\& \bar\eta
\end{array}
\right)=0\,,
\ee  
where the term $\gamma_{ab}$  with $a,b=\{\pi, h, \bar{\eta}\}$ corresponds to the coefficient of $b$ in  equation $a$ and they can be easily deduced from the above equations. Finally we set the determinant to zero and get
\be \label{Eq:DispRel}
k^4\omega(\omega+i\hub)\left[\omega^2+i 2 \hub \omega-\frac{(\lambda -1) \xi  ( 2 \xi-\eta )}{\eta  (3 \lambda -1)}k^2-\frac{\xi}{\eta}  \left( (\dot{\hub}-\hub^2) (\eta -3 \lambda +2 \xi +1)+(6 \lambda -4 \xi -2) \hub^2\right)\right]=0\;,
\ee
which can be written in a compact  form as 
\be \label{eq:disp22}
k^4\omega\l(\omega+i\f{\alpha}{2}\r)\left[\omega^2+i\alpha\omega-k^2c_s^2+\beta\right]=0\;.
\ee
From the above equation we deduce that  only one extra dynamical DoF exists, which corresponds to the scalar graviton ($\pi$ field in EFT language), as expected. Furthermore, one can identify the terms in the squared bracket as follows: $\alpha$ is a friction term, $\beta$ is the dispersion coefficient and  $c_s^2$ can be identified with the canonical speed of sound defined in vacuum, when no friction or dispersive terms are present. Let us notice that both the friction and dispersive terms are related to the nature of the dark energy component through the dependence of the Hubble rate on the latter~(\ref{friedmann1}). The procedure to compute the dispersion relation~(\ref{Eq:DispRel}) follows the one in ref.~\cite{Bloomfield:2012ff}, but here we include also  friction and dispersive terms.

In order to ensure that a given theory is viable, we enforce a set of  physical and mathematical viability conditions. The mathematical conditions prevent exponential instabilities from showing up in the solution of the $\pi$-field equation, and the physical ones correspond to the absence of ghosts and gradient instabilities for both scalar and tensor modes. In particular, in our analysis of Ho\v rava gravity, for the scalar DoF they correspond to 
\ba
 \frac{2 m_0^2\eta (1-3 \lambda ) k^2 }{(\eta -2 \xi ) \left(2(3 \lambda -1) \hub^2+\eta  (\lambda -1) k^2\right)}>0\,, \qquad \f{\xi(2\xi-\eta)(\lambda-1)}{\eta(3\lambda-1)}>0,
\ea
where the first condition corresponds to a positive kinetic term and it has been obtained from the action by integrating out all the non dynamical fields, while the second one ensures that the speed of sound is positive. Let us note that the ghost condition reduces to the one in the Minkowski background by setting the limit $a \rightarrow 1$.

Additional conditions to be imposed comes from the equation for the propagation of tensor modes $h_{ij}$,
\be
A_T(\tau) \ddot{h}_{ij}+B_T(\tau)\dot{h}_{ij} + D_T(\tau )k^2h_{ij} + E_{Tij} = 0.
\ee
where  $\delta T_{ij}$ generally contains the matter contributions coming from the neutrino and photon components and, for Ho\v rava gravity, the remaining coefficients read:
\begin{align}\label{tensorcoef}
&A_T=\f{2}{2\xi-\eta}\,, \hspace{1cm} B_T=\f{4\hub}{2\xi-\eta}\,,\\
&D_T= \f{2\xi}{2\xi-\eta}\,, \hspace{1cm} E_{Tij}=\f{a^2}{m_0^2}\delta T_{ij}\,.
\end{align}
The viability conditions require $A_T>0$ and $D_T>0$ to prevent respectively a tensorial ghost and gradient instabilities~\cite{Hu:2014oga}. 

It is easy to show that the above conditions translate into the following constraints on the parameters of Ho\v rava gravity:
\be
0<\eta<2\xi\,, \qquad \lambda>1\, \qquad \mbox{or}  \qquad  \lambda<\f{1}{3}, 
\ee
which are compatible with  the viable regions identified around a Minkowski background~\cite{Blas:2009qj}. 
In the following we will not explore the $\lambda<1/3$ branch since along it the cosmological gravitational constant on the FLRW background becomes unacceptably negative~\cite{Nariai:1973eg,Gurovich:1979xg} and the branch does not have a continuous limit to GR. 
The conditions that we have discussed are naturally handled by EFTCAMB/EFTCosmoMC in the form of \textit{viability priors} that are automatically enforced when the parameter space is being sampled.

Besides the above theoretical viability conditions, there are observational constraints on the Ho\v rava gravity parameters coming from existing data. 
In particular:
\begin{itemize}
\item Big Bang Nucleosynthesis (BBN) constraints~\cite{Carroll:2004ai}, which set an upper bound on $|G_{cosmo}/G_N-1|<0.38$ (99.7\% C.L.)\footnote{The original bound in ref.~\cite{Carroll:2004ai} is reported at 68\% C.L. and we convert it to 99.7\% C.L. by assuming a Gaussian posterior distribution of $G_{cosmo}/G_N-1$.}, where $G_{cosmo}$ is the cosmological gravitational constant as defined in section~\ref{Sec:background};
\item Solar system constraints, where the parametrized post Newtonian parameters (PPN) are bounded to be\footnote{The original bounds in ref.~\cite{Bell:1995jz} (and references therein) are reported at 90\% C.L. and we convert it to 99.7\% C.L. by assuming a Gaussian posterior distribution of the relevant parameters.}:  
\ba\label{PPN}
\alpha_1 < 3.0 \cdot 10^{-4} \,\, (99.7\% {\rm C.L.})  \,, \quad \alpha_2 < 7.0 \cdot 10^{-7} \,\, (99.7\% {\rm C.L.})  \,.
\ea 
where $\alpha_1$ and $\alpha_2$ are two of the parameters appearing in the PPN expansion of the metric around  Minkowski spacetime,  more precisely those  associated with the preferred frame effects~\cite{Will:2014kxa,Bell:1995jz}. 
Here we consider only these two parameters since they are the only ones of relevance for constraining LV.
It has been shown in refs.~\cite{Blas:2010hb,Blas:2011zd,Bonetti:2015oda}, that the PPN parameters for the low-energy action of Ho\v rava gravity, read
\ba
&&\alpha_1=4(2\xi-\eta-2)\,, \nn \\
&&\alpha_2=-\frac{(\eta -2 \xi +2) (\eta  (2 \lambda -1)+\lambda  (3-4 \xi )+2 \xi -1)}{(\lambda -1) (\eta -2 \xi )} \,.\label{PPNHorava}
\ea
 It is easy to show that combining the above relations, the above mentioned PPN bounds result in a direct constraint on $\lambda$ that reads:
\ba \label{Eq:PPNLambda}
&& {\rm log}_{10}\left( \lambda -1 \right) < -4.1  \,\, (99.7\% {\rm C.L.}) \,,
\ea
while the bound on $\alpha_1$ provides a degenerate constraint on the other two parameters $\{\xi, \eta\}$. 

\item $\check{\text{C}}$herenkov constraints from the observation of high-energy cosmic rays~\cite{Elliott:2005va} are usually imposed as a lower bound on the propagation speed of the scalar DoF and the propagation speed of tensor modes. In the case of LV theories we will refer the reader to Refs.~\cite{Elliott:2005va, Moore:2001bv}, for further details. However, since these bounds have not been worked out specifically for Ho\v rava gravity we decided not to impose them \textit{a priori}.
\end{itemize}
For the present analysis we consider two specific cases of Ho\v rava gravity:
\begin{enumerate}
\item Ho\v rava 3, hereafter H3, where we vary all three parameters $\{\lambda,\eta,\xi\}$ appearing in the low-energy Ho\v rava gravity action;
\item Ho\v rava 2, hereafter H2, where we choose the theory parameters in order to evade the PPN constraints~(\ref{PPNHorava}) by setting exactly $\alpha_1=\alpha_2=0$. This implies:  
\be\label{PPNHoravaconstraint}
\eta = 2\xi-2,
\ee
so that the number of free parameters reduce to two, $\{\lambda,\eta\}$. 
This case has the quality of systematically evading solar system PPN constraints,  meaning that it is not possible to build a local experiment, with arbitrary precision, to distinguish it from GR. Therefore it can only be constrained with cosmological observations. 
\end{enumerate}

For both cases we impose the physical and mathematical viability conditions in the form of \textit{viability priors} as discussed in ref.~\cite{Raveri:2014cka}. 
The portion of the  parameter space excluded by the \emph{viability priors} can be seen as a dark grey contour in figure~\ref{Fig:Horava_3D} for the H3 case and in figure~\ref{Fig:Horava_2D} for the H2 case. For both cases we also derive the bounds on $G_{cosmo}/G_N-1$ and for the H3 case we  provide cosmological bounds on the PPN parameters. These results are shown and discussed in detail in section~\ref{Sec:CosmologicalConstraints}.


\section{Ho\v rava Cosmology}\label{Sec:HoravaCosmology}
In this section we  highlight the cosmological implications of the low-energy Ho\v rava gravity cases, H2 and H3, previously introduced. 
In section~\ref{Sec:background} we  discuss the changes that Ho\v rava gravity induces at the level of the cosmological background, while in section~\ref{Sec:perturbations} we  elaborate on the effects that are displayed by the theory at the level of perturbations by means of two examples.

\subsection{Background}\label{Sec:background}
The first step towards testing a theory against cosmological observations, is to investigate the behaviour of its cosmological background.
In this section, we discuss the background evolution equation for Ho\v rava gravity, its  implementation in EFTCAMB, and review the definitions that we adopt for  the cosmological parameters.

The Ho\v rava gravity field equations for a flat FLRW background read:
\ba
&&\f{3\lambda-1}{2}\hub^2=\f{8\pi G_N(2\xi-\eta)}{6}a^2\sum_i\rho_{i}+\xi\f{\bar{\Lambda}}{3}a^2,\label{friedmann1}\\
&&-\f{3\lambda-1}{2}\l[\dot{\hub}+\f{1}{2}\hub^2\r]=-\f{\xi\bar{\Lambda}}{2}a^2+4\pi G_N \f{(2\xi-\eta)}{2} a^2 \sum_ip_{i},\label{friedmann2}
\ea
where  $\rho_{i}$ and $p_{i}$ are respectively the density and the pressure of the matter fluid components, i.e.  baryons and dark matter (m),  radiation and massless neutrino (r) and massive neutrinos ($\nu$). In this work we consider that all massive neutrino species have the same mass and we set the sum of their masses to be 0.06 eV. In addition to the Friedmann equations, we have the standard continuity equations for  matter and radiation:
\be\label{m_cons}
\dot{\rho}_i+3\hub(1+w_i)\rho_i=0,
\ee
while for massive neutrinos we refer the reader to ref.~\cite{Hu:2014sea} for a detailed discussion. 

Starting from the Friedmann eq.~(\ref{friedmann1}), we can define the cosmological gravitational constant as:
\be
G_{\rm cosmo}=\f{(2\xi-\eta)}{3\lambda-1}G_N,\label{eq:G_cosmo}
\ee
where it is clear that $G_{\rm cosmo}$ differs from $G_N$, which is obtained with local experiments, as already pointed out in ref.~\cite{Blas:2009qj}.
This definition allows us to write the Friedmann equation~(\ref{friedmann1}) in another way:
\be
\hub^2=8\pi G_{\rm cosmo}a^2\l(\f{\sum_i\rho_{i}}{3}+\f{1}{8\pi G_N}\f{2\xi}{2\xi-\eta}\f{\bar{\Lambda}}{3}\r)\,.
\ee
From this equation it is straightforward to see that in general, once the theory parameters have been properly set, the modification that Ho\v rava gravity induces at the level of the background is a global rescaling of $\hub$~\cite{Audren:2014hza}. 

In order to properly identify the parameters that we should fit to data,  we have to pay special attention to the working definition of all  the relevant quantities. 
In particular in the  definition of the relative density abundance. For the matter fields, we define $\Omega_{i}(a)$ in terms of the locally measured gravitational constant, $G_N$, and the present time Hubble parameter, $H_0$. We then derive the abundance of the effective dark energy, describing the modifications to the Friedmann equations, by means of the flatness condition, i.e.  $\sum_i\Omega_{i}(a)+\Omega_{DE}(a)=1$. To this extent, we rewrite the Friedmann eq.~(\ref{friedmann1}) as
\be \label{eq:FRWreshuffled}
\hub^2= 8\pi G_N \f{\sum_i \rho_{i}}{3}a^2+\f{2\xi}{2\xi-\eta}\f{\bar{\Lambda}}{3}a^2 +\l(1-\f{3\lambda-1}{2\xi-\eta}\r)\hub^2\,,
\ee
so that it is straightforward to identify
\ba
&&\Omega_{i}(a)=8\pi G_N \f{\rho_{i}}{3}\f{a^2}{\hub^2}\,,\label{matterdensity} \nn \\
&&\Omega_{DE}(a)=\f{2\xi}{2\xi-\eta}\f{\bar{\Lambda}}{3}\f{a^2}{\hub^2}+1-\f{3\lambda-1}{2\xi-\eta}\label{DEdensity}.
\ea
At present time ($a_0=1$), we can immediately see that $\Omega_{\rm DE}^0 = 1-\sum_i\Omega_{i}^0$ with:
\be\label{DEtoday}
\Omega_{DE}^0=\f{2\xi}{2\xi-\eta}\f{\bar{\Lambda}}{3H_0^2}+1-\f{3\lambda-1}{2\xi-\eta}.
\ee
This allows us to rewrite the Friedmann eq.~(\ref{friedmann1}) in terms of the parameters that we are going to sample as:
\be\label{Hubble}
\hub^2=\f{(2\xi-\eta)}{3\lambda-1}a^2 H_0^2 \l[ \f{\Omega_m^0}{a^3}+\f{\Omega_r^0}{a^4}+\rho_{\nu} +\l(\Omega_{DE}^0-1+ \f{3\lambda-1}{2\xi-\eta}\r)\r]\,.
\ee
This is the background equation that EFTCAMB evolves, along with its time derivatives. For details about how the code treats $\rho_{\nu}$ see ref.~\cite{Hu:2014sea}. Finally, one can use eq.~(\ref{DEtoday}), to substitute the ``bare'' cosmological constant with  $\Omega_{DE}^0$, therefore in the following we use the latter as one of the Ho\v rava parameters that we fit to data instead of $\bar{\Lambda}$.
\begin{figure}[t!]
\centering
\includegraphics[width=1\textwidth]{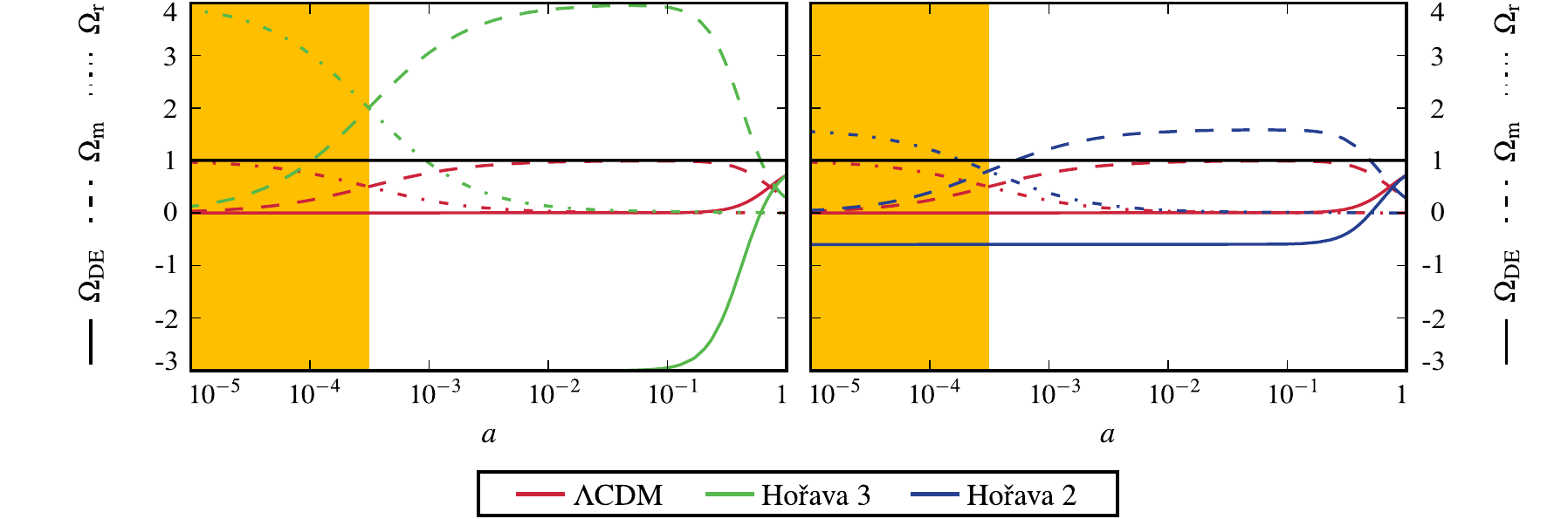}
\caption{The figure shows the evolution of the densities parameters for baryons and dark matter (m, dashed line), radiation, neutrino and massive neutrinos (r, dot dashed line) and dark energy (DE, solid line).   In the left panel we compare the density parameters of the H3 case (green lines) with the ones in the $\Lambda$CDM model (red lines). In the right panel the comparison is between the H2 case (blue lines) and $\Lambda$CDM.  The yellow area highlights the radiation dominated era. For this figure the standard cosmological parameters are chosen to be $\Omega_b^0\,h^2=0.0226$, $\Omega_c^0\,h^2=0.112$, $\Omega_{\nu}^0\,h^2=0.00064$ and $H_0=70 \,\mbox{Km}/\mbox{s}/\mbox{Mpc}$.  In the H3 case the Ho\v rava gravity parameters are $\lambda = 1.4$, $\xi = 0.9$, $\eta = 1.0$ while in the H2 case they are fixed to $\lambda = 1.4$, $\eta = 1.0$.}
\label{Fig:Background}
\end{figure}

We shall now specialize to some choices of the Ho\v rava parameters, and derive  the corresponding expansion history in order to visualize and discuss the effects of Ho\v rava gravity, in particular for the H3 and H2 cases, on background cosmology.  We choose the background values of the cosmological parameters to be $\Omega_b^0\,h^2=0.0226$ for baryons, $\Omega_c^0\,h^2=0.112$ for cold dark matter, $\Omega_{\nu}^0\,h^2=0.00064$ for massive neutrinos and $H_0=70 \,\mbox{Km}/\mbox{s}/\mbox{Mpc}$, accordingly to the default CAMB parameters. Additionally, the parameters of the H3 case are chosen to be: $\lambda = 1.4$, $\xi = 0.9$, $\eta = 1.0$; while in the H2 case we set $\lambda = 1.4$ and $\eta = 1.0$. While the general trend of the modifications does not depend on the magnitude of the theory parameters,  the above values are selected in order to enhance the effects and clearly display the changes with respect to the standard cosmological model, $\Lambda$CDM. Thus they have to be considered as illustrative examples because the values involved are significantly bigger than the observational bounds that we will derive in section~\ref{Sec:CosmologicalConstraints}. However, in both cases the choices of parameters respect the viability criteria discussed in section~\ref{Stability}. 

In figure~\ref{Fig:Background} we can see the behaviour of the relative densities for matter (dark matter and baryons),   radiation (photons and relativistic neutrinos),  and effective dark energy, as defined in eqs.~(\ref{matterdensity}). One can notice that at early times the matter species display density values that are generally bigger than one, on the contrary the dark energy component assumes  negative values.  This can be explained as follows. The matter components are well behaved, with positively defined densities with a time evolution that is exactly the standard one (eq.~(\ref{m_cons})), as expected when working in  Jordan frame. However, the expansion history changes as it is rescaled by a constant (eq.~(\ref{Hubble})), altering the time behaviour of the relative abundances. The effective dark energy balances this effect in order to respect the flatness condition. We argue that in this specific case the interpretation of the modification of gravity  in terms of a fluid-like component is not well justified/posed, representing instead  a genuine geometrical modification of the gravitational sector. This kind of behaviour for the effective dark energy component is commonly encountered in dynamical analysis studies of modified gravity models where the flatness condition is used as a constraint equation~\cite{Amendola:2006we,Frusciante:2013zop}.  
From figure~\ref{Fig:Background}, we can also notice that Ho\v rava gravity does not affect the time of radiation-matter equality as the continuity equations for these species are not changed, as it is clearly highlighted by the yellow region in the figure. Indeed $\Omega_m$ and $\Omega_r$ for all the models cross at the same value of the scale factor. On the other hand, the time of equality between matter and dark energy is slightly modified depending on the model parameters. 
Finally, let us notice that, once the parameters of the theory are chosen to be compatible with the observational constraints, all these effects that we have discussed are quite mitigated and become hardly noticeable by eye in the plots. Indeed, values of the parameters consistent with the bounds that we derive in section~\ref{Sec:CosmologicalConstraints} would induce a  less negative DE density at earlier times. 
%
\subsection{Perturbations} \label{Sec:perturbations}

In this section, we proceed to study the dynamics of cosmological perturbations.  Once we  have worked out the background equations of Ho\v rava gravity~(\ref{Hubble}), as well as the  mapping of this theory into the EFT language~(\ref{Horava_mapping}), we have all the ingredients required by EFTCAMB to perform an accurate analysis of the perturbations. For technical details on the actual implementation, as well as the full set of perturbative equations that are evolved by EFTCAMB, we refer the reader to ref.~\cite{Hu:2014oga}.

As we will see, the behaviour of  perturbations in Ho\v rava gravity displays an interesting and rich phenomenology, allowing to investigate the theory and to constrain its parameters with the available data.
In the following, we perform an in depth analysis of the dynamics of linear perturbations and the corresponding observables, specializing to a choice of parameters for the case H3 and one for the case H2, in order to visualize and quantify the modifications.  In all cases,  we set the values of the cosmological parameters to the one used in the previous section, which are the default CAMB parameters, while for the Ho\v rava parameters we use:  in the H3 case, $(\xi-1)=-0.01$, $(\lambda-1)=0.004$, $\eta=0.01$; in the H2 case, $(\lambda-1)=0.02$, $\eta=0.05$. 
As it will be clear in the next section, these are noticeably bigger than the observational constraints that we will derive, but they facilitate the visualization of the effects on the observables.  Let us stress that, while the direction and entity of the modifications that will be described in the remaining of this section are specific to the choice of parameters, we have found an analogous trend for several choices of parameters that we have sampled in the region allowed by the \emph{viability priors}.

\begin{figure}[!t]
\centering
\includegraphics[width=1\textwidth]{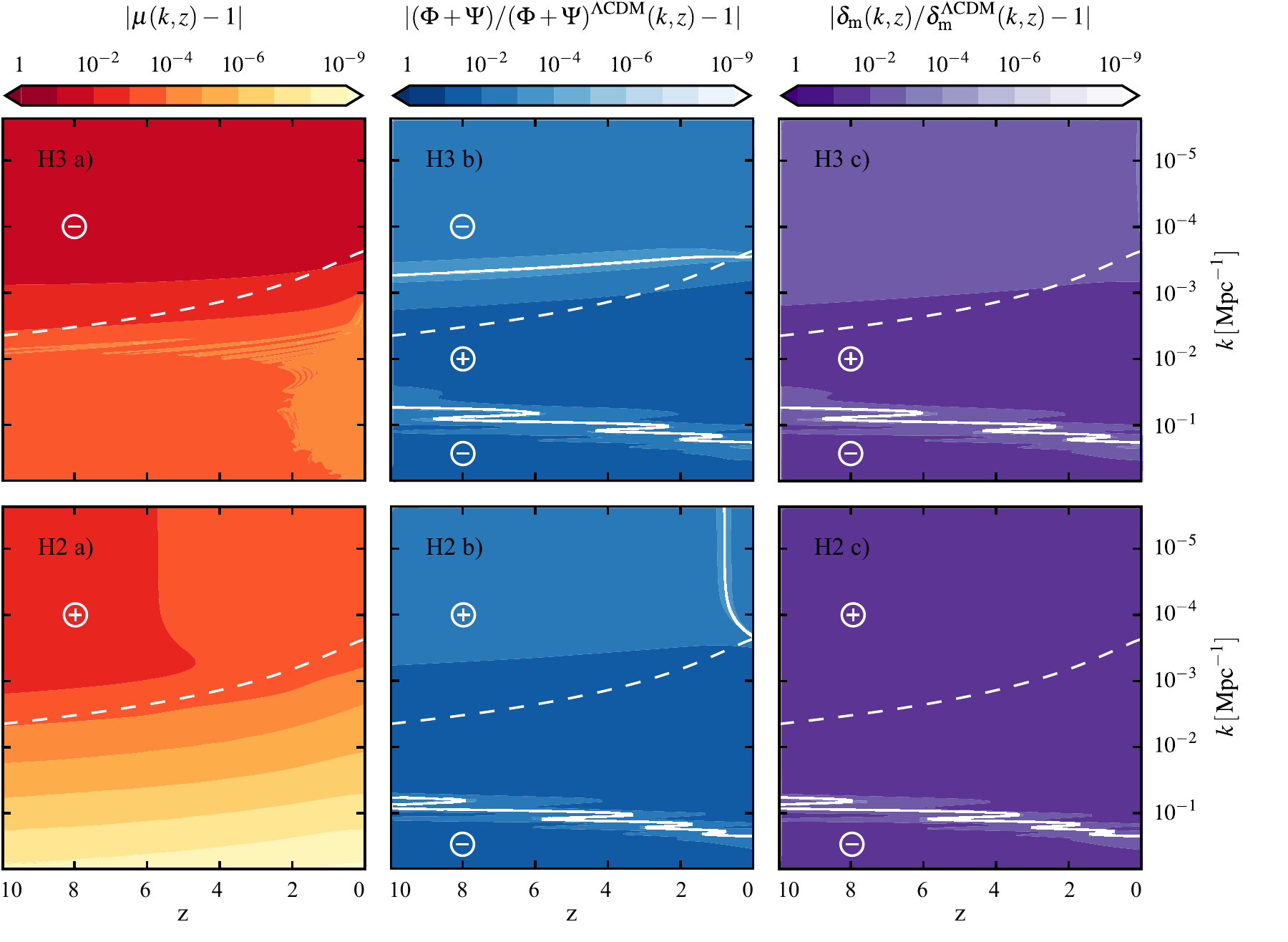}
\caption{ We show the relative comparison of the modification of the Poisson equation $\mu$, the source of gravitational lensing  $\Phi+\Psi$ (whose derivative sources the ISW effect on the CMB), and $\delta_{\rm m}\equiv \sum_m \rho_{\rm m} \Delta_{\rm m}/ \sum_m \rho_{\rm m}$ with their $\Lambda$CDM values for the H3 (upper panel) and H2 (lower panel) models. In all panels, the dashed white line represents the physical horizon while the solid white line shows where the relative comparison changes sign. 
For this figure the standard cosmological parameters are chosen to be $\Omega_b^0\,h^2=0.0226$, $\Omega_c^0\,h^2=0.112$, $\Omega_{\nu}^0\,h^2=0.00064$ and $H_0=70 \,\mbox{Km}/\mbox{s}/\mbox{Mpc}$.
 In the H3 case the additional parameters are $(\xi-1)=-0.01$, $(\lambda-1)=0.004$, $\eta=0.01$ while in the H2 case they are fixed to $(\lambda-1)=0.02$, $\eta=0.05$. For a detailed explanation of this figure see section~\ref{Sec:perturbations}. }
\label{Fig:Mu_Gamma}
\end{figure}
\begin{figure}[!t]
\centering
\includegraphics[width=1\textwidth]{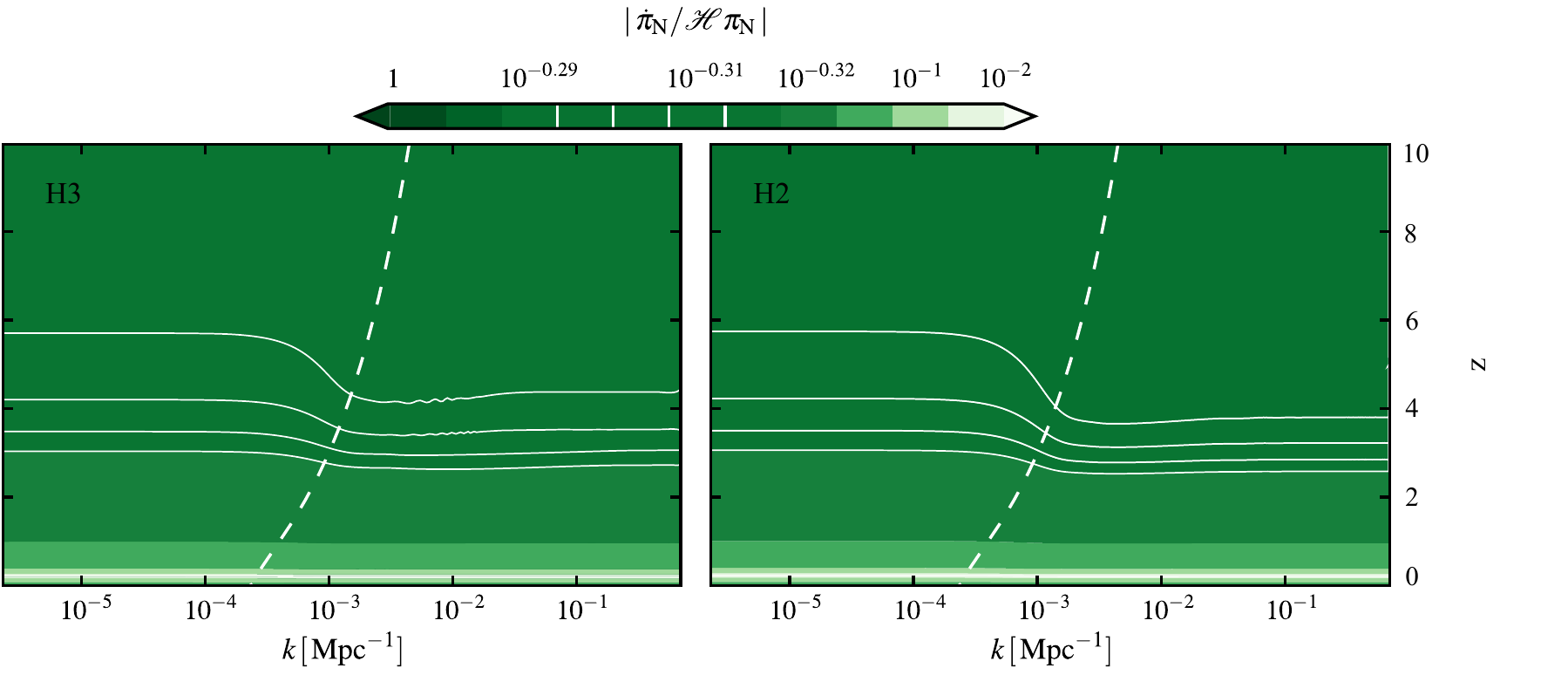}
\caption{ We show the quantity $\xi_N= \dot{\pi}_N / \hub\pi_N$ that we introduce as an indicator of the goodness of the quasi-static approximation for the H3 (left panel) and H2 (right panel) cases. In both panels, the dashed white line represents the physical horizon,  while the solid white lines highlight the scale dependence of this quantity.
For this figure the standard cosmological parameters are chosen to be $\Omega_b^0\,h^2=0.0226$, $\Omega_c^0\,h^2=0.112$, $\Omega_{\nu}^0\,h^2=0.00064$ and $H_0=70 \,\mbox{Km}/\mbox{s}/\mbox{Mpc}$. In the H3 case the Ho\v rava gravity
 parameters are $(\xi-1)=-0.01$, $(\lambda-1)=0.004$, $\eta=0.01$ while in the H2 case they are fixed to $(\lambda-1)=0.02$, $\eta=0.05$. For a detailed explanation of this figure see section~\ref{Sec:perturbations}.}
\label{Fig:QS}
\end{figure}

Let us now focus on the time and scale evolution of cosmological perturbations and the growth of structure. In order to discuss the deviations of  Ho\v rava gravity from $\Lambda$CDM, we study the behaviour of the $\mu(k,a)$-function, which is defined in  Newtonian gauge as~\cite{Silvestri:2013ne}
\be\label{mugamma}
k^2\Psi\equiv-\mu(k,a)\f{a^2}{2m_0^2}\rho_m \Delta_m \,,
\ee 
where $\Delta_m$ is the comoving matter density contrast and $\Psi$ is the scalar perturbation describing fluctuations in  the time-time component of the metric. As it is clear from eq.~(\ref{mugamma}), $\mu$ parametrizes deviations from GR in the Poisson equation. In the standard cosmological model, $\Lambda$CDM,  this function is constant  and $\mu=1$. 
Let us notice that  EFTCAMB does never evolve the above quantity~(\ref{mugamma}), but it can easily output $\mu$ as a derived quantity.
Moreover, we also analyse the behaviour of the quantity $\Phi+\Psi$, where $\Phi$ is the scalar perturbation of the space-space component of the metric in Newtonian gauge. This quantity is important as it  allows to identify possible modifications in the lensing potential and in the low multipole of the cosmic microwave background (CMB) radiation through the Integrated Sachs-Wolfe (ISW) effect. Finally, we explore the fluctuations in the total  matter distribution defined as $\delta_{\rm m}\equiv \sum_m \rho_{\rm m} \Delta_{\rm m}/ \sum_m \rho_{\rm m}$. 

In figure~\ref{Fig:Mu_Gamma} we show the time and scale behaviour of these three quantities. In order to facilitate the visualization of the deviations from the $\Lambda$CDM behaviour, we show the logarithmic fractional comparison between these quantities in the two Ho\v rava gravity cases considered and the $\Lambda$CDM model. 

\begin{figure}[!th]
\centering
\includegraphics[width=1\textwidth]{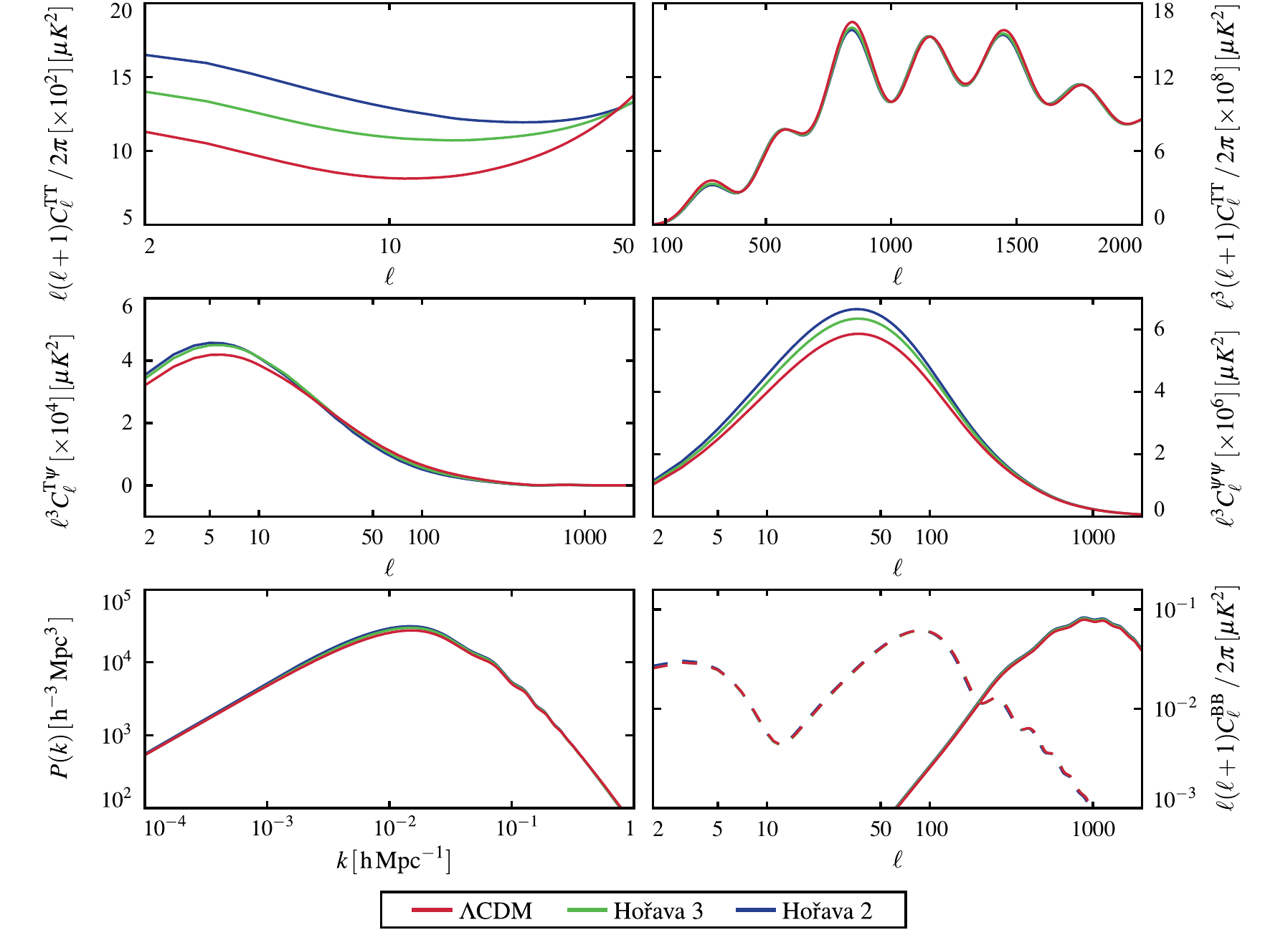}
\caption{Power spectra of different cosmological observables in the $\Lambda$CDM, H2 and H3 cases.
{\it Upper panel}: CMB temperature-temperature power spectrum at large (left) and small (right) angular scales.
{\it Central panel}: lensing potential and CMB temperature cross correlation power spectrum (left), lensing potential auto correlation power spectrum (right).
{\it Lower panel}: matter power spectrum (left) and B-mode polarization power spectrum (right). In this last panel the solid line corresponds to the scalar induced B-mode signal while the dashed one shows the tensor induced component.
For this figure the standard cosmological parameters are chosen to be $\Omega_b^0\,h^2=0.0226$, $\Omega_c^0\,h^2=0.112$, $\Omega_{\nu}^0\,h^2=0.00064$ and $H_0=70 \,\mbox{Km}/\mbox{s}/\mbox{Mpc}$. The Ho\v rava gravity parameters in
 H3 case are chosen to be: $(\xi-1)=-0.01$, $(\lambda-1)=0.004$, $\eta=0.01$; in the H2 case they are: $(\lambda-1)=0.02$, $\eta=0.05$. }
\label{Fig:Spectra}
\end{figure}
\begin{itemize}
\item {\it H3 case}: from the top left panel of figure~\ref{Fig:Mu_Gamma} we can see that $\mu$ significantly deviates from one at large scales and all redshift with fractional differences that are  around unity (100\%). Small deviations of the order of $10^{-4}$ can be also seen at  small scales and high redshift.
At small scales and low redshift, in the bottom right part of the H3 a) panel, one can notice small features due to the fact that the $\pi$ field oscillates while being coupled to the other species.
From the top central panel of the same figure we can see that gravitational lensing is modified as well. On large, super-horizon, scales deviations from the $\Lambda$CDM behaviour are not significant, staying below $10^{-2}$ at all the times shown.  In general at these scales the lensing is suppressed. On sub-horizon scales in turn the  enhancement of the lensing potential with respect to the $\Lambda$CDM case becomes relevant.  A similar behaviour can be seen in the total matter density contrast.  Although on super-horizon scales, as well as just below the horizon, the density contrast is enhanced compared to the $\Lambda$CDM one, on very small scales it is suppressed. Noticeably the oscillations that we see in $\mu$ do not reflect on $\Phi+\Psi$ and $\delta_m$, which  look rather regular. The physical interpretation of this is that even if the additional scalar DoF is introducing fluctuations in the structure of the Poisson equation the field is not coupled strongly enough to introduce fast fluctuations in the matter and metric fields themselves.
\item {\it H2 case}: from the lower left panel of figure~\ref{Fig:Mu_Gamma} we can notice that, in the H2 case, the behaviour of $\mu$ is rather different from the H3 case. In particular on small scales its value returns to the GR one. This is compatible with the extra constraint that we have imposed in this case~(\ref{PPNHoravaconstraint}), making the theory indistinguishable from GR on solar system scales.
On large scales and high redshift, similarly to the H3 case, deviations from the $\Lambda$CDM behaviour are of the order $10^{-2}$ (1\%).
Panels H2 b) and H2 c) in figure~\ref{Fig:Mu_Gamma} show that the lensing effects and the growth of matter perturbations do not follow the trend of $\mu$. Indeed, in the case of  lensing, in panel H2 b), around and below the horizon the model displays significant deviations from the $\Lambda$CDM behaviour that are similar to the H3 ones.
From panel H2 c) we notice that the growth of matter perturbations deviates significantly from the $\Lambda$CDM one (around $10^{-2}$) at almost all redshifts and scales. Finally, in the same panel it can be noticed that the density contrast is enhanced  for $k\lesssim 10^{-1} $ h/Mpc while it is suppressed at very small scales and all redshift. 
\end{itemize}

After considering the cosmological evolution of metric and matter perturbations we now turn to the study of the dynamics of the additional scalar DoF that propagates in Ho\v rava gravity. In particular we study the quantity introduced in ref.~\cite{Hu:2013twa} to quantify the deviations from quasi-staticity for the dynamical scalar DoF, $\pi$, i.e.
\be \label{eq:QS}
\xi_N=\f{\dot{\pi}_N}{\hub\pi_N}\,,
\ee
where with the index $N$ we indicate that we are working with the $\pi$-field in Newtonian gauge. 
This quantity compares the evolutionary time-scale of the additional scalar DoF with the Hubble time scale, thus quantifying how many times the $\pi$-field changes significantly in a Hubble time. Small values of this quantity imply that the $\pi$ field is slowly evolving and that time derivatives of the field can be neglected when compared to the value of the scalar field itself. On the contrary large values mean that the time derivative of the field is playing a major dynamical role, and hence QS would not be a safe assumption.

The time and scale behaviour of $\xi_N$ can be  seen, for the H3 and H2 cases, in figure~\ref{Fig:QS}.  We can notice that, roughly for both cases, the $\pi$-field is slowly evolving at low redshift ($0<z<1$), on the other hand, at higher redshift we can see that its dynamics becomes relevant and deviations from a QS behavior are order 30\%.
We can also notice that, at all scales and times, the evolutionary time scale of $\pi_{N}$ is smaller than the Hubble rate. 
From the same figure we can see that this evolutionary rate does not significantly depend on scale. The white lines in figure~\ref{Fig:QS} show some residual scale dependence at early times and clearly show that this scale dependence gets weaker at late times.

\begin{figure}[!t]
\centering
\includegraphics[width=1\textwidth]{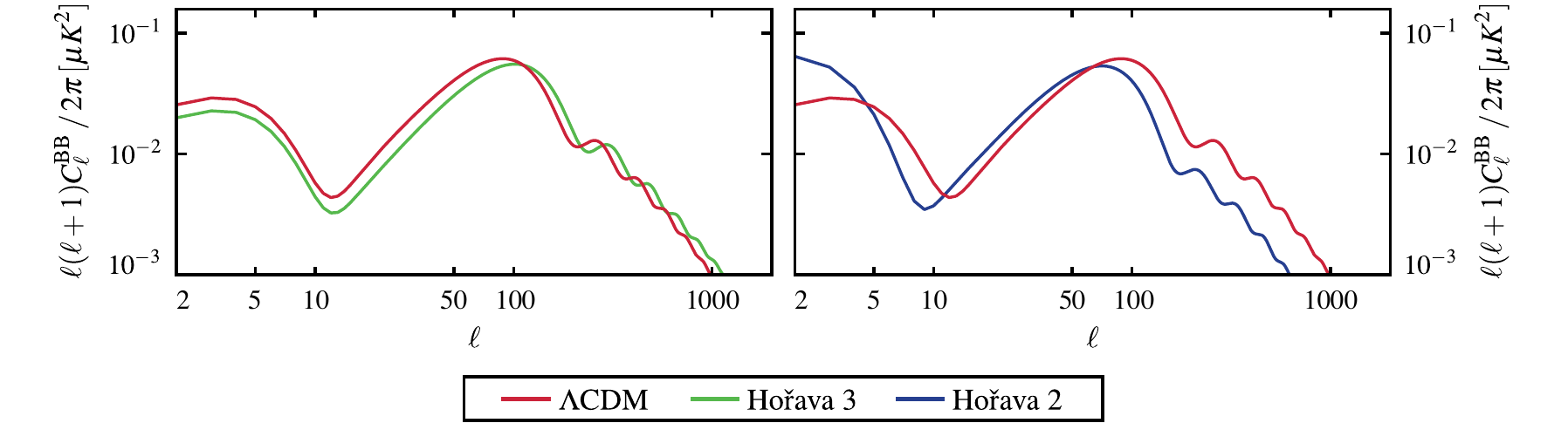}
\caption{The tensor induced component of the B-mode CMB polarization power spectrum in the $\Lambda$CDM, H2 and H3 cases.
For this figure the standard cosmological parameters are chosen to be $\Omega_b^0\,h^2=0.0226$, $\Omega_c^0\,h^2=0.112$, $\Omega_{\nu}^0\,h^2=0.00064$, $r=1$ and $H_0=70 \,\mbox{Km}/\mbox{s}/\mbox{Mpc}$. The Ho\v rava gravity parameters in H3 case are chosen to be: $(\xi-1)=-0.3, (\lambda-1)=4\times10^{-4},\eta=10^{-3}$ in the H2 case they are: $(\lambda-1)=1,\eta=0.6$. }
\label{Fig:SpectraBB}
\end{figure}

Finally, we discuss how the modified dynamics of perturbations in Ho\v rava gravity  affects the observables that we later use to constrain this theory.
In figure~\ref{Fig:Spectra}, we compare several power spectra for the H2 and H3 cases in comparison to the $\Lambda$CDM model.  We identify the following effects on the observables: 
\begin{itemize}
\item \textit{Differences in the  late time Integrated Sachs-Wolfe (ISW) effect}. For the two cases that we explore, we find an enhancement of the amplitude of the low-$\ell$ temperature power spectrum,  as it can be seen from the top left panel of figure~\ref{Fig:Spectra} which is related to an increase of the late-time ISW effect~\cite{Sachs:1967er}. The latter  is sourced by the time derivative of $\Phi+\Psi$ and, as we can see from figure~\ref{Fig:Mu_Gamma}, for the two Ho\v rava gravity cases the time evolution of this 
quantity is modified. This change also affects the CMB temperature-lensing cross correlation (central left panel), as discussed below.

\item \textit{Differences in the gravitational lensing}.  As we already discussed, in the specific cases that we explore, gravitational lensing results to be  enhanced as we can see in the central panel of  figure~\ref{Fig:Mu_Gamma}. This reflects on the CMB lensing power spectrum as shown in the central right panel of figure~\ref{Fig:Spectra}, where we can notice that fluctuations of this  observable are enhanced for both H3 and H2 cases with respect to the $\Lambda$CDM model. This modification also has an effect on the high multipole of the lensed CMB temperature power spectrum as highlighted in the top-right panel of figure~\ref{Fig:Spectra}. At first glance we can see that, compared to the $\Lambda$CDM model, the profile of the high-$\ell$ peaks is  less sharper in the H3 and H2 cases because of the lensing  enhancement.  We can also notice that there is a slight asymmetry between peaks and troughs  due to a combined effect of the lensing modification with the modified  Hubble rate discussed in section~\ref{Sec:background}, thus leading to a small change in the angular scale of the CMB peaks.  From the central left panel, we can see that the CMB temperature-lensing cross correlation spectrum is influenced by  both the ISW and lensing modifications. In particular, this spectrum results to be enhanced at low-$\ell$  because of the lensing and ISW enhancements but it is suppressed for $50<\ell<100$  following the trend of the temperature power spectrum. Indeed, we can notice, from the top right panel of figure~\ref{Fig:Spectra}, that at these scales the spectra are suppressed due to the lensing effect as previously mentioned.
Finally, the enhancement of the lensing potential  also affects the component of the CMB B-mode power spectrum that is sourced by the lensing of the E-mode of polarization. This situation is highlighted  in the lower right panel of figure~\ref{Fig:Spectra}. The solid lines representing this component of the B-mode spectrum  are enhanced proportionally to the  enhancement  in the lensing potential. 
\item  \textit{Differences in the growth of matter perturbations and the distribution of the large scale structure.} 
 For the two cases under analysis (H3 and H2),  we observe a slight enhancement of the growth of structure in the total matter power spectrum, at intermediate scales, as well as a slight suppression on small scales, as it is clearly depicted in the lower left panel of figure~\ref{Fig:Spectra}, and in agreement with our previous analysis of the density contrast, see figure~\ref{Fig:Mu_Gamma}. 
The matter power spectrum, for both H2 and H3, follows the $\Lambda$CDM one on  large scales ($k \lesssim  10^{-3}$ h/Mpc) while  for $ 10^{-3} \lesssim k \lesssim 10^{-1}  $ h/Mpc it is slightly enhanced, particularly for the  H2 case. At very small scales, both the H2 and H3 matter spectra follow the $\Lambda$CDM  behaviour. 
\item \textit{Differences in the propagation of tensor modes.} As previously discussed in section~\ref{Stability}, the tensor dynamical equation is modified in Ho\v rava gravity.  This change is usually reflected in the tensor induced component of the B-modes of CMB polarization~\cite{Amendola:2014wma,Raveri:2014eea}. In particular, in the H3 case, the parameter $\xi$ controls directly the propagation speed of gravitational waves, while the combination $2\xi-\eta$ is responsible for the strength of coupling between tensor modes and matter. Instead, in the H2 case the tensor speed of sound is controlled by $\eta$, while there is no effect on the coupling with matter. The  choice of parameters we made for figure~\ref{Fig:Spectra}, displays a significant effect on the scalar component of the B-mode spectrum as shown in the lower right panel of figure~\ref{Fig:Spectra} as solid lines, but the effect on the tensor component (dashed line) of the B-mode power spectrum for the same parameters is much smaller and not visible in the figure.  
In figure~\ref{Fig:SpectraBB} we change the Ho\v rava gravity parameters to better display the effect of the change in the tensor sector. Therefore only for this figure we choose the Ho\v rava gravity parameters in H3 case to be $(\xi-1)=-0.3, (\lambda-1)=4\times10^{-4},\eta=10^{-3}$ and  in the H2 case they are $(\lambda-1)=1,\eta=0.6$.  As we can see from that figure, the leading effect is due to the modification of the speed of gravitational waves~\cite{Raveri:2014eea}. In the next section we will find that due to a combination of viability requirements and data constraints,  for the H3 case, $\xi\leq 1$, therefore the spectrum results to be shifted to the right with respect to the $\Lambda$CDM one, since tensor modes propagate sub-luminally. On the other hand, in the H2 case, tensor modes propagate super-luminally ($\eta>0$) and the whole spectrum is shifted to the left. Finally, a modification of the coupling to matter leaves an observational imprint that is much smaller that the previous one as cosmological gravitational waves propagate almost in vacuum. We can also conclude, on the basis of the results we will present in the next section, that in the H3 case since the  tensor sound speed is less than one, the $\check{\text{C}}$herenkov constraints are  not always satisfied but only in a very tiny range~\cite{Elliott:2005va,Moore:2001bv}. On the contrary, in the H2 case the tensor sound speed is always super-luminal, then the $\check{\text{C}}$herenkov constraints are evaded.

\end{itemize}

\section{Cosmological constraints}\label{Sec:CosmologicalConstraints}

In this section we derive and discuss the observational constraints on Ho\v rava gravity coming from cosmological probes. After describing the data sets used, we focus on the H3 and H2 cases described at the end of section~\ref{Stability}.

\subsection{Data sets}\label{Sec:dataset}

In our analysis  we  use several geometrical and dynamical probes, combining them progressively. 

The first data set employed, hereafter PLC, consists of the low-$\ell$ ($2\leq\ell<50$) CMB temperature-temperature power spectra  from the \emph{Planck} satellite~\cite{Ade:2013kta,Ade:2013zuv}, considering the 9 frequency channels ranging from $30\sim353$ GHz.
In addition, we consider the $100$, $143$, and $217$ GHz frequency channels for the high-$\ell$ modes ($50\leq\ell\leq2500$) of the CMB temperature spectrum.
We also include the WMAP low-$\ell$ polarization spectra ($2\leq\ell\leq32$)~\cite{Hinshaw:2012aka} to break the degeneracy between the re-ionization optical depth and the amplitude of CMB temperature anisotropy.

The second data set considered, denoted as BG, is a combination of background measurements that helps to break the degeneracies between background parameters and the ones describing the behaviour of perturbations. We  use data from HST~\cite{Riess:2011yx} which measures the local Hubble constant from optical and infrared observations of more than 600 Cepheid variables. In addition, we  consider the ``Joint Light-curve Analysis'' (JLA) Supernovae sample as analysed in ref.~\cite{Betoule:2014frx} which is constructed from the SNLS, SDSS and HST SNe data, together with several low redshift SNe. We also employ baryon acoustic oscillations measurements taken from: the SDSS Main Galaxy Sample at $z_{\rm eff}=0.15$~\cite{Ross:2014qpa}; the BOSS DR11 ``LOWZ" sample at $z_{\rm eff}=0.32$~\cite{Anderson:2013zyy}; the BOSS DR11 CMASS at $z_{\rm eff}=0.57$ of~\cite{Anderson:2013zyy}; and the 6dFGS survey at $z_{\rm eff}=0.106$~\cite{Beutler:2011hx}. 

The third data set that we use, consists of the \emph{Planck} 2013 full-sky lensing potential power spectrum obtained by using the $100$, $143$, and $217$ GHz frequency bands employed by the \emph{Planck} collaboration to detect the CMB lensing signal with a significance greater than $25\sigma$~\cite{Ade:2013tyw}.
We  refer to this data set as the lensing one, hereafter CMBL. 

Finally, we use the measurements of the galaxy power spectrum by the WiggleZ Dark Energy Survey~\cite{wigz} in order to exploit the constraining power of large-scale structure data.
The WiggleZ data set consists of the galaxy power spectrum measured from spectroscopic redshifts of $170,352$ blue emission line galaxies over a volume of $1\,\mbox{Gpc}^3$~\cite{Drinkwater:2009sd,Parkinson:2012vd}. The covariance matrices are taken to be the ones given in ref.~\cite{Parkinson:2012vd} and are computed using the method described in ref.~\cite{Blake:2010xz}. It has been shown in refs.~\cite{Parkinson:2012vd,Dossett:2014oia} that linear theory predictions are a good fit to the data regardless of non-linear corrections up to a scale of $k\sim 0.2\,\mbox{h}/\mbox{Mpc}$. Since changes in the growth induced by modifications of gravity can slightly alter this scale, in this work, we use the WiggleZ galaxy power spectrum with a more conservative cut of $k_{\rm max} = 0.1\,\mbox{h}/\mbox{Mpc}$. We marginalize over a scale independent linear galaxy bias for each of the four redshift bins, as in ref.~\cite{Parkinson:2012vd}. 
Let us notice that, in general, the linear galaxy bias in a modified gravity scenario is not scale independent, as shown in refs.~\cite{Hui:2007zh,Parfrey:2010uy}. 
However, as we will show in the next section, the cosmological constraints are mainly driven by CMB and  background observables towards the $\Lambda$CDM limit. Therefore, we do not expect that a scale dependent bias will dominate the results in Ho\v rava gravity when considering WiggleZ data.

\subsection{H3 case: results}\label{Sec:H3results}

\begin{figure}[!t]
\centering
\includegraphics[width=1\textwidth]{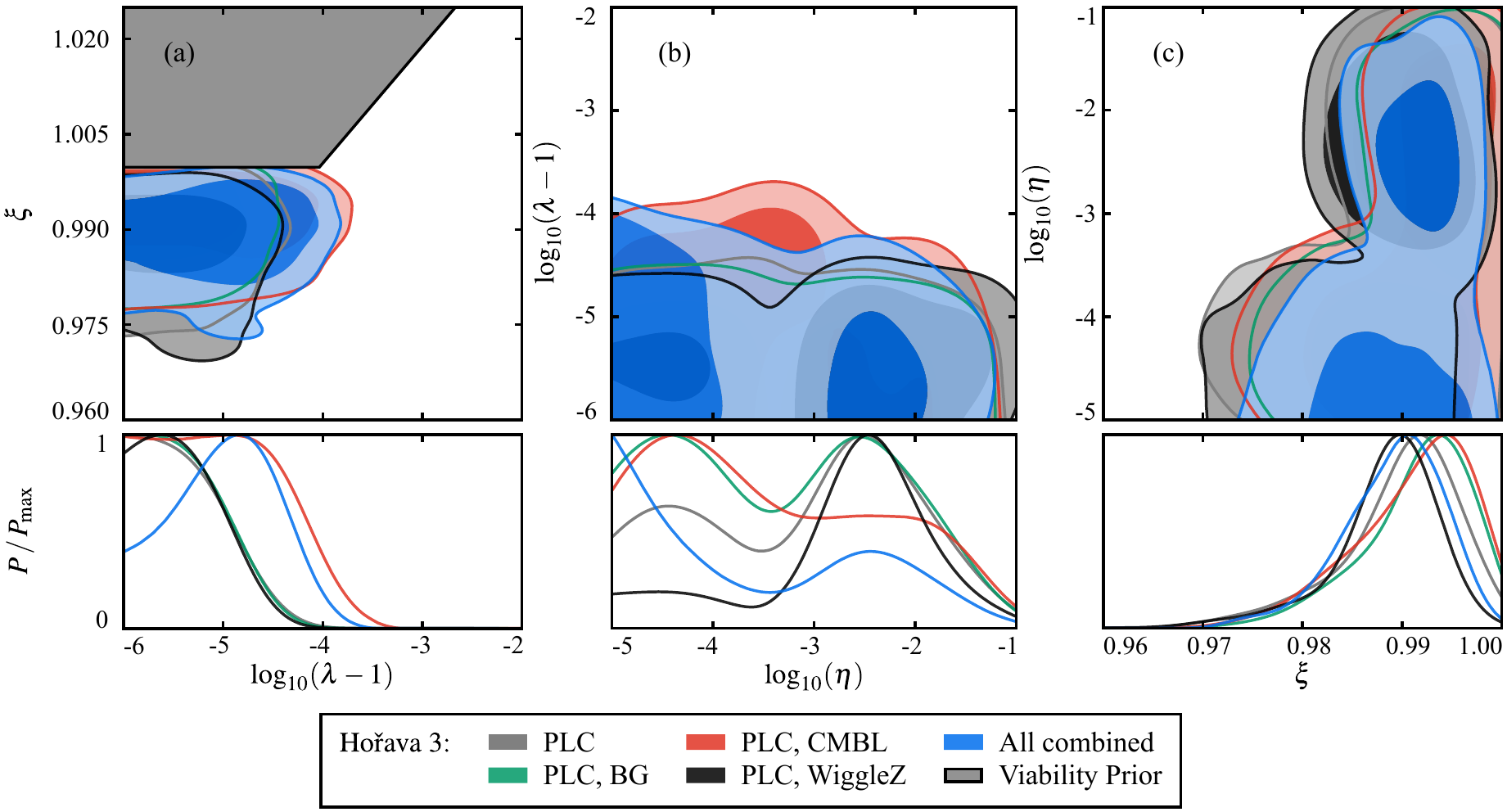}
\caption{Results of our analysis of the H3 case. {\it Upper panel}: The marginalized joint likelihood for combinations of the parameters of low-energy Ho\v rava gravity. The darker and lighter shades correspond respectively to the $68\%$ C.L. and the $95\%$ C.L..
{\it Lower panel}: The marginalized likelihood of the parameters of low-energy Ho\v rava gravity.
In both panels different colours correspond to different data set combinations as shown in legend. The dark grey shade corresponds to the marginalized region of parameter space excluded by \emph{viability priors}.}
\label{Fig:Horava_3D}
\end{figure}

\begin{table}[t!]
\centering
\begin{tabular}{|c|c|c|c|c|c|c|}
\hline
\multicolumn{1}{|c|}{  }&
\multicolumn{6}{c|}{H3 case}\\
\hline
\hline
Parameters & Prior & PLC & PLC+BG & PLC+CMBL & PLC+WiggleZ & all combined \\
\hline
\hline
$\xi-1$ & $[-0.1,0.1]$ & $-0.01^{+0.01}_{-0.02}$ & $-0.01^{+0.01}_{-0.02}$ & $-0.01^{+0.01}_{-0.02}$ & $-0.01^{+0.01}_{-0.02}$ & $-0.01^{+0.01}_{-0.02}$ \\[1mm]
$\log_{10} ({\lambda}-1)$ & $[-8,-2]$ & $<-4.56$ & $<-4.68$ & $<-4.24$ & $<-4.68$ & $<-4.31$ \\[1mm]
$ \log_{10} \eta$ & $[-5,-1]$ & $--$ & $--$ & $--$ & $--$ & $--$ \\[1mm]
$\Omega_{DE}^0$ & derived & $0.69_{-0.07}^{+0.06}$ & $0.69\pm 0.02$ & $0.70_{-0.05}^{+0.04}$ & $0.66\pm 0.06$ & $0.69\pm 0.02$ \\[1mm]
\hline
\hline
$ \alpha_1 $ & derived & $<0.283$ & $<0.240$ & $<0.263$ & $<0.322$ & $<0.220$\\[1mm]
$ \log_{10} \alpha_2 $ & derived & $<5.05$ & $<4.83$ & $<5.04$ & $2.70^{+2.52}_{-2.41}$ & $<4.72$\\[1mm]
$ G_{\rm cosmo}/G_N -1 $ & derived & $<0.035$ & $<0.030$ & $<0.033$ & $<0.040$ & $<0.028$\\[1mm]
\hline
\end{tabular}
\caption{The $99.7\%$ C.L. marginalized posterior bounds on the H3 case parameters and relevant derived parameters.}
\label{Tab:Constraints_Horava_3D_99}
\end{table}

The first case  we compare to cosmological observations is the  low-energy limit of Ho\v rava gravity, H3, for which the parameters of the theory are $\{\xi,\lambda,\eta\}$.

We  sample $\lambda$ and $\xi$ shifting them by one so that the GR limit of the new parameters, i.e. $\lambda-1$ and $\xi-1$, corresponds to a zero value.
In addition, we  use a logarithmic sampler for the parameters $\eta$ and $\lambda-1$ as they are constrained to be positive by physical viability, as discussed  at length in section~\ref{Stability}. Instead, $\xi$ is allowed to change sign, hence we sample $\xi-1$ linearly.

When combining the \emph{viability  priors} discussed in section~\ref{Stability}, with cosmological data, we notice that the requirement of physical viability has a strong effect on the posterior distribution of the parameter $\xi$. This is shown in panel (a) of figure~\ref{Fig:Horava_3D}. Even though $\xi$ is not constrained to be above or below $1$ a priori, the magnitude of  $\lambda$, draws the posterior of the model into a region where only values of $\xi \leq 1$ are viable. Let us notice that there is a very small viable region in the parameter space above one ($1<\xi\lesssim 10^{-6}$), which due to the scale adopted in the plot cannot be seen by eye; however, since this is significantly smaller than the region below $\xi=1$, it is not picked up when sampling linearly the parameter $\xi$.
As we can see from the other two panels of  figure~\ref{Fig:Horava_3D}, the \emph{viability priors} do not have a strong impact on $\lambda$ and $\eta$  in the region where the posterior of the model is peaked. 

From the top panels of figure~\ref{Fig:Horava_3D}, we can  notice that the different parameters specifying Ho\v rava gravity do not have sizable degeneracies between them over all the range explored (which spans several orders of magnitude). 
 From the lower panel of the same figure we can notice that different data sets contribute differently to the parameter bounds.
In particular we can see that PLC strongly constrains the $\lambda$ parameter, while preferring a bigger value for $\eta$ and $\xi$. The addition of background probes pushes these two parameters closer to the $\Lambda$CDM limit of the theory. Noticeably the addition of CMB lensing strongly degrades the bounds on $\lambda$ while being consistent with PLC+BG for the other two parameters. This behavior is expected considering the known tension between the \emph{Planck} 2013 and the \emph{Planck} 2015 data and the CMB lensing power spectrum as reconstructed from the CMB trispectrum~\cite{Ade:2013zuv,Ade:2015xua,Hu:2015rva}.
These results are confirmed by the marginalized bounds on the H3 parameters reported in table~\ref{Tab:Constraints_Horava_3D_99}.
In particular the $99.7\%$ C.L. bounds on the $\xi$ and $\eta$ parameters weakly depend on the data set used and in particular for the $\eta$ parameter no  $99.7\%$ C.L. bounds are found. 
As discussed in section~\ref{Sec:background}, the ``bare'' cosmological constant $\bar{\Lambda}$ has been replaced in this analysis by $\Omega_{DE}^0$, and the latter has been included in table~\ref{Tab:Constraints_Horava_3D_99}. The most relevant result that can be drawn from  table~\ref{Tab:Constraints_Horava_3D_99} is that low-energy Ho\v rava modifications of gravity are severely constrained, with the corresponding parameters bounded to be orders of magnitude below unity. In particular, we find that cosmological data have a strong constraining power on $\lambda$. Here we summarize the bounds, at $99.7\%$ C.L, that  we get from the combination of all data sets:
\ba
&&\xi-1 = -0.01^{+0.01}_{-0.02}\,, \nonumber\\
&&\mbox{log}_{10}(\lambda-1)<-4.31\,,\nonumber\\
&&\Omega_{DE}^0=0.69\pm 0.02 
\ea 

Let us notice that the \emph{viability priors} give also an upper bound on $\xi$, i.e. $\xi\leq 1$.  All the marginalized constraints on standard and derived cosmological parameters for the H3 case are shown in appendix~\ref{appCP}.
 
In table~\ref{Tab:Constraints_Horava_3D_99} we report also the marginal bounds on the  PPN parameters  $\{\alpha_1,\alpha_2\}$ and $G_{\rm cosmo}/G_N-1$. From the first two, we can notice the extreme complementarity of cosmological and solar system experiments in constraining Ho\v rava gravity. The cosmological observations lead to an upper bound on $\alpha_1$ that is $\alpha_1<0.220$, while PPN bounds on this parameter are three orders of magnitude stronger. 
Cosmological observations are, however, sensitive to $\xi$ and weakly sensitive to $\eta$, while solar system probes constrain just a degenerate combination of the two.
On the other hand the cosmological bounds on the  parameter $\alpha_2$ are worst than the PPN ones by several orders of magnitude. That is due to the fact that in the limit in which $\lambda$ is constrained to be  smaller than the other parameters by cosmological data, the $\alpha_2$ parameter goes to infinity as it is clear from its definition~(\ref{PPNHorava}). 
If we compare the cosmological constraint on $\lambda$ with the one that is derived from $\alpha_2$ in eq.~(\ref{Eq:PPNLambda}), we see that our bounds are compatible with the solar system constraints.
Finally, we can see that the bound on  $ G_{\rm cosmo}/G_N-1<0.028\, (99.7\% \,\mbox{C.L., all combined})$ is improved by one order of magnitude with respect to previous results~\cite{Carroll:2004ai}.

\subsection{H2 case: results}\label{Sec:H2results}
\begin{figure}[!t]
\centering
\includegraphics[width=1\textwidth]{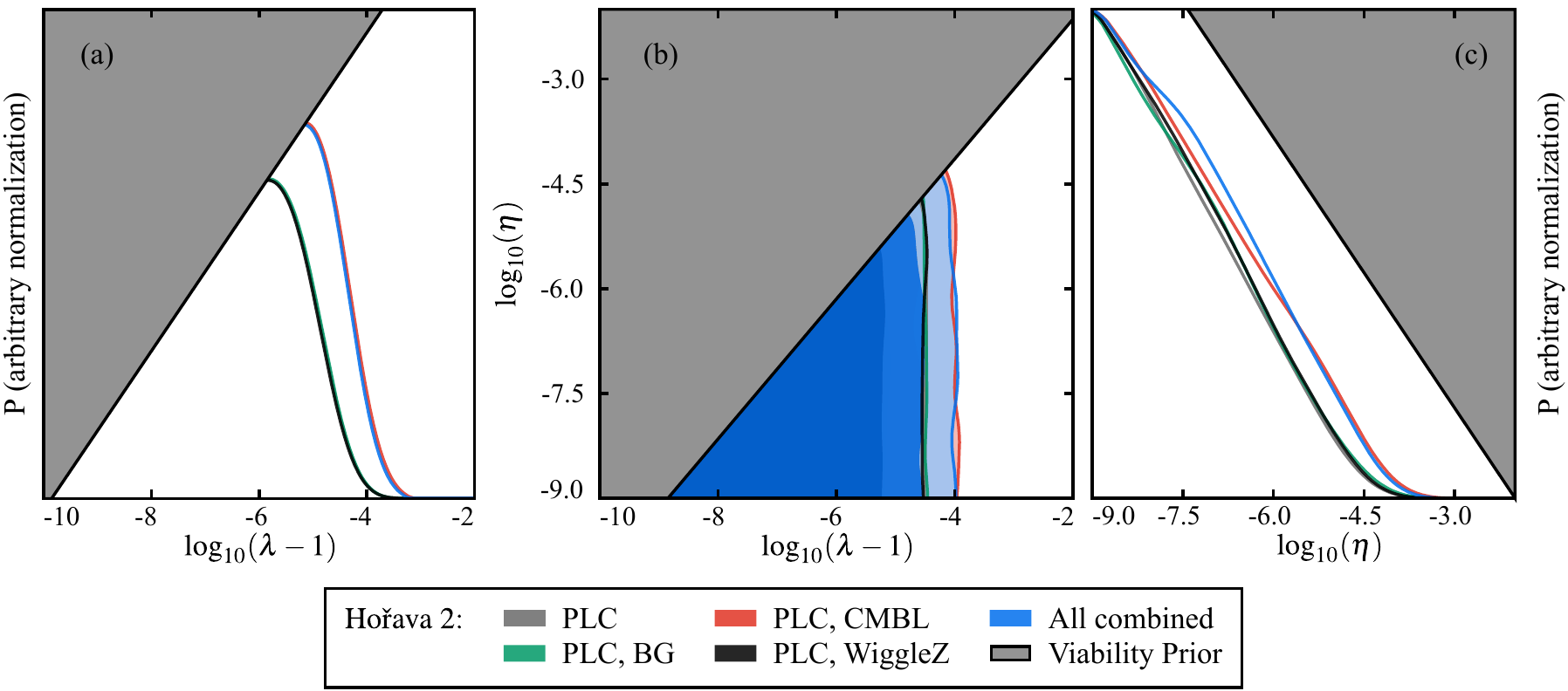}
\caption{Results of our analysis of the H2 case. {\it Panel (a)}: The marginalized likelihood of ${\rm log}_{10}\left(\lambda -1 \right)$; {\it Panel (b)}: The marginalized joint likelihood of ${\rm log}_{10}\left(\lambda -1 \right)$ and ${\rm log}_{10}\eta$. The darker and lighter shades correspond respectively to the $68\%$ C.L. and the $95\%$ C.L.. The theoretical viability condition is roughly $(\lambda-1) > \mathcal{O}(2\eta/9)$; {\it Panel (c)}: The marginalized likelihood of ${\rm log}_{10}\eta$. In all panels different colours correspond to different data set combinations as shown in legend. The dark grey shade corresponds to the marginalized region of parameter space excluded by \emph{viability priors}. The arbitrary normalization of the posterior is chosen to better display the effect of the \emph{viability priors}.}
\label{Fig:Horava_2D}
\end{figure}
The second case  we consider, is a sub-case of the previous one obtained by restricting to the plane of the parameter space corresponding to $\eta = 2\xi -2$. For this choice,  the solar system constraints are automatically evaded as shown by studies of the PPN limit of Ho\v rava gravity (see section~\ref{Stability}). We  refer to this as the H2 case.
The free parameters of the theory are now  $\{\eta,\lambda\}$ and, as discussed in the previous section, we  sample the parameter space of the logarithm of $\eta$ and $\lambda-1$ since both these quantities are constrained by the \emph{viability priors} to be positive. Unlike for the H3 case, where the \emph{viability priors} had a strong influence only on the parameter $\xi$, in the H2 case they have a strong influence on the marginalized posterior for both $\eta$ and $\lambda$, as one can see in figure~\ref{Fig:Horava_2D}. 
In particular one can notice in panel (b) of figure~\ref{Fig:Horava_2D} that the viable region is a triangle corresponding roughly to the condition $(\lambda-1) > \mathcal{O}(2\eta/9)$. This triangle shape of the marginalized joint posterior of the model parameters has a strong influence on the marginalized 1D posterior of the two parameters.
In particular, as we can see in panel (a) of figure~\ref{Fig:Horava_2D} , the low tail of $\lambda$ is cut by the \emph{viability priors} and, in panel (c), the posterior of $\eta$ becomes markedly non-gaussian. Apart from the degeneracy induced by this prior cut, no other degeneracy between the Ho\v rava gravity parameters is present.

In panel (a) of figure~\ref{Fig:Horava_2D},  we can see that different data sets contribute differently to the cosmological bounds. In particular, the PLC, PLC+BG and PLC+WiggleZ  data sets are pushing the posterior of $\lambda$ to smaller values. The constraints become slightly weaker when considering CMB lensing for the same reasons explained in the previous section and it dominates  the bounds coming from the total data set combination. Because of the degeneracy induced by the \emph{viability priors},  in the H2 case we are able to set  bounds also on  $\eta$.  These are shown in table~\ref{Tab:Constraints_Horava_2D_99}. In this table we can notice that the $99.7\%$ C.L. confidence bounds on $\lambda$ are comparable to the previous ones and the bounds on $\eta$ are considerably stronger. Indeed the bounds for all combination of data sets for the H2 case are:
\ba
&&\mbox{log}_{10}(\lambda-1)<-4.39 \,,\nonumber\\
&&\mbox{log}_{10}(\eta)< -4.51 \,, \nonumber\\
&&\Omega_{DE}^0=0.69\pm 0.02 \,.
\ea 
As in the previous case $\Omega_{DE}^0$ and all the standard derived cosmological parameters  are reported in appendix~\ref{appCP}.
 
Noticeably in the H2 case, the bounds on $G_{\rm cosmo}/G_N-1 (< 6.1 \times 10^{-5}, \,99.7\%\,\mbox{C.L., all combined}$) are more stringent than in the H3 case. In particular this bound is several orders of magnitude stronger than the BBN bound. 

 The H2 case studied in this section has been already investigated in ref.~\cite{Audren:2014hza}. Indeed, our H2 case is the khronometric model in ref.~\cite{Audren:2014hza} when the LV in matter is switched off and a precise mapping between the different notations adopted is worked out. In this respect, a comparison between our results and the one in ref.~\cite{Audren:2014hza} is not straightforward. We find an overall good agreement in the cosmological predictions by comparing the output of the code used by~\cite{Audren:2014hza} to EFTCAMB. 
The stronger bound on $G_{\rm cosmo}/G_N-1$ is then due to the different choices of cosmological parameters that we sample and the different priors that we impose on them. 
\begin{table*}[t!]
\centering
\begin{tabular}{|c|c|c|c|c|c|c|}
\hline
\multicolumn{1}{|c|}{  }&
\multicolumn{6}{c|}{H2 case}\\
\hline
\hline
Parameters & Prior & PLC & PLC+BG & PLC+CMBL & PLC+WiggleZ & all combined \\
\hline
\hline
$ \log_{10} ({\lambda}-1)$ & $[-10, -2]$ & $<-4.62$ & $<-4.62$ & $<-4.38$ & $<-4.59$ & $<-4.39$ \\[1mm]
$ \log_{10} \eta$ &  $[-10, -2]$ & $<-4.68$ & $<-4.58$ & $<-4.40$ & $<-4.73$ & $< -4.51$ \\[1mm]
$\Omega_{DE}^0$ & derived & $0.68_{-0.06}^{+0.04}$ & $0.69\pm 0.02$ & $0.69\pm 0.04$ & $0.66_{-0.06}^{+0.05}$ & $0.69\pm 0.02$ \\[1mm]
\hline
\hline
$ G_{\rm cosmo}/G_N -1 $ & derived  &  $<3.6\times10^{-5}$ & $<3.6\times10^{-5}$ & $<6.3\times10^{-5}$ & $<3.9\times10^{-5}$ & $<6.1\times10^{-5}$ \\[1mm]
\hline
\end{tabular}
\caption{ The $99.7\%$ C.L. marginalized posterior bounds on the H2 case parameters and $G_{\rm cosmo}/G_N-1$.}
\label{Tab:Constraints_Horava_2D_99}
\end{table*}
%

\section{Conclusion}\label{conclusion}
In this paper we have performed a thorough investigation of the cosmology of Ho\v rava gravity, which is a Lorentz violating theory  proposed as a candidate for quantum gravity~\cite{Horava:2008ih,Horava:2009uw}. The emergence of Lorentz violations at all scales, makes the theory power-counting renormalizable at very high energies ($\sim 10^{19}$ GeV/c$^2$). However, since at low energies we do not experience LV effects, we expect large scales tests to place important constraints on the theory. In particular, we have analysed Ho\v rava gravity at cosmological scales, to see whether there is any room for LV at these energies. As we will summarize in details in the following, we have found that cosmological data severely constrain Ho\v rava gravity.

We have performed our analysis within the EFT framework for dark energy and modified gravity~\cite{Gubitosi:2012hu,Bloomfield:2012ff}, which we have reviewed in section~\ref{eft}. We have focused on the dynamics of the background and linear perturbations, and considered the power-counting renormalizable action for Ho\v rava gravity which includes all the \emph{quadratic} operators with up to sixth spatial derivatives~\cite{Blas:2009qj}. For this action, we have worked out a complete mapping into the EFT language, in  section~\ref{mapping}, finding that its low-energy part  is completely mapped in the most commonly used EFT action~(\ref{actioneft}). While the high-energy part requires the inclusion of additional operators, that we have identified in appendix~\ref{L4L6}.  After working out the full mapping, when exploring the cosmology and corresponding observational bounds, in sections~\ref{Sec:HoravaCosmology} and~\ref{Sec:CosmologicalConstraints}, we have restricted to the low-energy part of the action which is  sufficient for a first exploration of  the large scale phenomenology of the theory.  

For our analysis we have made use of the powerful EFTCAMB/EFTCosmoMC package~\cite{Hu:2013twa,Raveri:2014cka,Hu:2014oga}. While this package was made publicly available by some of the authors at \url{http://wwwhome.lorentz.leidenuniv.nl/~hu/codes/}, for our analysis we have used an updated version which fully implements Ho\v rava gravity. By the latter we mean that, for the first time, we have implemented in EFTCAMB a \emph{full mapping} case,  solving the Ho\v rava equations for the background, instead of using a \textit{designer} approach to the expansion history.  We have included a detailed discussion of this procedure  in section~\ref{Sec:background}. After solving the background and before proceeding to the evolution of the perturbations, EFTCAMB runs a check on the viability of the selected theory, enforcing some stability requirements  such as the absence of ghosts and gradient instabilities. The latter conditions become \textit{viability priors} when using EFTCosmoMC to constrain the parameters of the theory by means of cosmological data. Finally, we have proceeded to evolve linear scalar perturbations with the general EFT equations~\cite{Hu:2014oga}, specializing their coefficients to the corresponding expressions in the Ho\v rava case through the mapping worked out in section~\ref{mapping}. 

Let us notice that besides the physical stability requirements, Lorentz violations can also be constrained via BBN and solar system tests, as discussed at length in section~\ref{Stability}. In our analysis we have not imposed the BBN experimental bounds a priori,  rather we have compared them to our finding for  cosmological constraints. 
Nevertheless, we have investigated two cases: a first one, H3, where the low-energy Ho\v rava gravity parameters $\{\lambda,\xi,\eta\}$ were allowed to vary freely; the second case, H2, where we enforced a relationship between the parameters that allows the theory to evade PPN constraints, reducing the number of free parameters to two, $\{\lambda,\eta\}$. 

In section~\ref{Sec:HoravaCosmology}, we have studied in details the cosmology of Ho\v rava gravity. At  background level, we have found a constant rescaling of the Hubble rate which reflects in the behaviour of the  density parameters. Indeed, as shown in figure~\ref{Fig:Background}, the fractional matter density exceeds unity at all times and the fractional density of the effective dark energy~(\ref{DEdensity})  correspondingly becomes negative, so that the flatness condition is satisfied at all times. This behaviour signals that the modifications of gravity in this theory should be considered as a purely geometrical effect, rather than be interpreted in terms of a dark fluid.
At the level of linear perturbations we have identified  modifications in the ISW effect, the gravitational lensing, the rate of growth of structure and the B-modes spectrum, which translated into stringent bounds on the Ho\v rava parameters when we fit them to cosmological data. In order to facilitate a in depth visualization  of  characteristic features of Ho\v rava gravity at the level of perturbations, we have specialized to two choices of parameters, one for the H3 case and one for the H2 case, and have performed a thorough analysis of the dynamics of perturbations and the corresponding cosmological observables.  In both cases, we have found a general  enhancement of the growth of matter perturbations and  the lensing potential. The first modifies the shape of the matter power spectrum, which we have found to be  enhanced for $ 10^{-3}<k<10^{-1}$ h/Mpc; the second one impacts  the CMB temperature power spectrum at high-$\ell$ and the CMB lensing power spectrum.  In particular in the latter case the effect is of a general  enhancement of power in the lensing potential auto spectrum.
A modification of the lensing potential also alters the scalar perturbation induced component of the B-mode power spectrum, leading to an  enhancement  in that signal.  On the other hand, we found an enhancement of  the CMB temperature power spectrum at large angular scales and its cross-correlation with the lensing potential. This effect can be traced back to an enhancement of the  ISW effect at late times. Finally, the tensor power spectrum is also modified as the speed of sound of the tensor modes depends on the parameters of the theory. In particular the tensor BB-power spectrum is shifted on the right with respect to the $\Lambda$CDM one for the H3 case as the tensor propagation is sub-luminal and on the left for H2 because the propagation is super-luminal.   Noticeably, we have determined that in general the quasi-static approximation is not safe to describe the evolution of sub-horizon perturbations in Ho\v rava gravity if we want an accuracy better than 30\%.
Let us stress that while the direction and entity of these modifications is dependent on the specific choice of parameters, we have found a general  enhancement of the growth rate, lensing potential, and an enhancement of the ISW effect for several choices of parameters that we have sampled in the region allowed by the \emph{viability priors}.

In section~\ref{Sec:CosmologicalConstraints}, we have moved on to perform a global fit of the two cases of low-energy Ho\v rava gravity, H3 and H2,  to progressive combinations of cosmological data sets: the CMB temperature-temperature and lensing power spectra by \emph{Planck} 2013,  WMAP low-$\ell$ polarization spectra, the WiggleZ galaxy power spectrum, the local Hubble measurements and Supernovae data from SNLS, SDSS and HST and the BAO measurements from BOSS, SDSS and 6dFGS. 

In the case of H3, we have set upper bounds on $\lambda$ and a lower bound on $\xi$, while for $\Omega_{DE}^{(0)}$ (which through eq.~(\ref{DEtoday}) replaces $\bar{\Lambda}$) we found a mean value and errors that are close to the ones of the $\Lambda$CDM model.  Specifically we obtained $\xi-1=-0.01^{+0.01}_{-0.02}$, $\mbox{log}_{10}(\lambda-1)<-4.31$ and $\Omega_{DE}^0=0.69\pm 0.02$ at $99.7\%$ C.L. for the  combination of all the data sets considered. For all the other data set combinations see table~\ref{Tab:Constraints_Horava_3D_99}. 
As a general result we have found that the values of the Ho\v rava gravity parameters are constrained to be close to their values in the GR limit. For both cases we reported the constraints on the standard cosmological parameters in tables~\ref{Tab:CosmologicalConstraintsH2}-\ref{Tab:CosmologicalConstraintsH3}. Moreover, for the H3 case we get an  improved bound on $G_{cosmo}/G_N-1 < 0.028$ ( $99.7\%$ C.L., for the combination of all data sets) which outruns the BBN one.  On the other hand PPN experiments are  three orders of magnitude stronger in constraining the $\alpha_1$ parameter that we find to be $\alpha_1<0.220$ ( $99.7\%$ C.L., for the combination of all data sets), while our cosmological bound on $\lambda$  is compatible with  the one derived from solar system tests. 

For the H2 case,  we were able to set upper bounds on $\mbox{log}_{10}(\lambda-1)<-4.39$ and $\mbox{log}_{10}(\eta)<-4.51$ and constraints on $\Omega_{DE}^0=0.69 \pm 0.02$ at $99.7\%$ C.L. with all data sets. Noticeably for this model we get a  quite stringent bound on $G_{cosmo}/G_N-1 < 6.1 \times 10^{-5} $ at $99.7\%$ C.L. by combining all the considered data sets.

The full mapping of the low-energy limit of Ho\v rava gravity has been publicly released as part of an update of EFTCAMB/EFTCosmoMC.
As part of future work, it would be certainly of interest to explore the phenomenology associated to the high-energy part of the Ho\v rava gravity action to see whether additional operators can affect significantly linear perturbations.  Future analysis could also include the study of LV in the dark matter sector.

\section*{Acknowledgements}
We are grateful to Carlo Baccigalupi, Enrico Barausse and  Thomas Sotiriou for useful discussions and comments on the manuscript. 
 We are indebted with Mikhail Ivanov,  Diego Blas and Sergey Sibiryakov for sharing their code that greatly helped us in cross checking some of the results of this work.
The research of NF and DV 
has received funding from the European Research Council under the European Community's Seventh Framework Programme (FP7/2007-2013, Grant Agreement No.~307934).
 MR acknowledges partial support from the INFN-INDARK initiative. BH is supported by the Dutch Foundation for Fundamental Research on Matter (FOM). AS acknowledges support from The Netherlands Organization for Scientific Research (NWO/OCW), and also from the D-ITP consortium, a program of the Netherlands Organisation for Scientific Research (NWO) that is funded by the Dutch Ministry of Education, Culture and Science (OCW).

\appendix
\section{The $L_4$ and $L_6$ Lagrangians}\label{L4L6}

The $L_4$ and $L_6$ Lagrangians contain, respectively, all the operators up to fourth and sixth order spatial derivatives, which are compatible with the symmetry of Ho\v rava gravity and guarantee its power-counting renormalizability. 
The number of those operators is very large, but given that we are interested in an effective field theory description of linear scalar perturbations, only the ones which are second order in perturbations have to be considered in the action.
The latter have been identified in ref.~\cite{Blas:2009qj} and they are given by suitably contracting the three-dimensional Ricci tensor and scalar, the acceleration $a_i$, and their spatial derivatives. In detail, the HE part of action~(\ref{horavaaction}) can be written as 
\ba\label{HEaction}
\mathcal{S}_{H,HE}&=&\f{1}{16\pi G_H}\int{}d^4x\sqrt{-g}\left(g_1\mathcal{R}^2+g_2\mathcal{R}_{ij}\mathcal{R}^{ij}+g_3\mathcal{R}\nabla_i a^i 
+g_4 a_i \nabla^2a^i \right. \nn\\ 
&&\left.+g_5\mathcal{R} \nabla^2 \mathcal{R}+g_6\nabla_i\mathcal{R}_{jk}\nabla^i\mathcal{R}^{jk}+g_7a_i\nabla^4 a^i+g_8\nabla^2\mathcal{R}\nabla_i a^i \right)
\ea
where $\nabla^2=\nabla^i\nabla_i$ and $\nabla^4=\nabla^i\nabla_i\nabla^j\nabla_j$, and the coefficients $g_i$ are running coupling constants of suitable dimensions. The  first and second lines contain respectively the contributions from $L_4$ and $L_6$. 

We expand now the above action up to second order in perturbations by considering that on a flat FRLW background the components of $\mathcal{R}$ and $\mathcal{R}_{ij}$ identically vanish. Then we map it into the language at the basis of the EFT formalism discussed in section~\ref{eft} by using the relation~(\ref{aiexpansion}) for $a_i$. With these recipes, it is straightforward to show that the operators in action~(\ref{HEaction}) up to second order can be written as 
\begin{subequations}\label{Hor_ops}
\begin{align}
&g_1\f{m_0^2}{(2\xi-\eta)}\mathcal{R}^2=  g_1\f{m_0^2}{(2\xi-\eta)}(\delta \mathcal{R})^2 ,  \\
&g_2\f{m_0^2}{(2\xi-\eta)}\mathcal{R}_{ij}\mathcal{R}^{ij} =  g_2\f{m_0^2}{(2\xi-\eta)}\delta \mathcal{R}_{ij}\delta \mathcal{R}^{ij}, \\
&g_3\f{m_0^2}{(2\xi-\eta)}\mathcal{R}\nabla_i a^i = g_3\f{m_0^2}{2(2\xi-\eta)}\delta \mathcal{R}\nabla^2(a^2\delta g^{00}),  \\
&g_4 \f{m_0^2}{(2\xi-\eta)}a_i \nabla^2a^i= g_4 \f{m_0^2}{4(2\xi-\eta)}  \partial_i (a^2 g^{00}) \nabla^2 \partial^i (a^2 g^{00}), \\
&g_5\f{m_0^2}{(2\xi-\eta)}\mathcal{R} \nabla^2 \mathcal{R} = g_5\f{m_0^2}{(2\xi-\eta)} \delta \mathcal{R} \nabla^2 \delta \mathcal{R}, \\
&g_6\f{m_0^2}{(2\xi-\eta)}\nabla_i\mathcal{R}_{jk}\nabla^i\mathcal{R}^{jk}= g_6\f{m_0^2}{(2\xi-\eta)}\nabla_i \delta \mathcal{R}_{jk}\nabla^i \delta \mathcal{R}^{jk},  \\
&g_7 \f{m_0^2}{(2\xi-\eta)}a_i\nabla^4 a^i=g_7\f{m_0^2}{4(2\xi-\eta)}\partial_i(a^2\delta  g^{00})\nabla^4 (\partial^i (a^2\delta g^{00})),  \\
&g_8 \f{m_0^2}{(2\xi-\eta)}\nabla^2\mathcal{R}\nabla_i a^i = g_8\f{m_0^2}{2(2\xi-\eta)}\nabla^2 \delta\mathcal{R}\nabla^2 (a^2\delta g^{00}).
\end{align}
\end{subequations}
We notice immediately that  the EFT action~(\ref{actioneft}) is incomplete if one wants to treat the full version of Ho\v rava gravity (which includes the operators in action~(\ref{HEaction})), then we need to add to it all the operators in eqs.~(\ref{Hor_ops}). This suggests to extend the EFT action discussed in section~\ref{eft} to a more general one by adding the following part:
\ba\label{EFTact_addendum}
\mathcal{S}_{EFT,2}&=&\int{}d^4x\sqrt{-g}\left[\lambda_1(\tau)(\delta \mathcal{R})^2+\lambda_2(\tau)\delta \mathcal{R}_{ij}\delta \mathcal{R}^{ij}+\lambda_3(\tau)\delta \mathcal{R}\nabla^2(a^2\delta g^{00})+\lambda_4(\tau)\partial_i (a^2 g^{00})\nabla^2 \partial^i (a^2 g^{00})\right. \\ &+&\left.\lambda_5(\tau)\delta \mathcal{R} \nabla^2 \delta \mathcal{R}
+\lambda_6(\tau)\nabla_i \delta \mathcal{R}_{jk}\nabla^i \delta \mathcal{R}^{jk}+\lambda_7(\tau)\partial_i(a^2\delta  g^{00})\nabla^4 (\partial^i (a^2\delta g^{00}))+\lambda_8(\tau)\nabla^2 \delta\mathcal{R}\nabla^2 (a^2\delta g^{00})\right].\nonumber
\ea
In the Ho\v rava gravity case the EFT functions $\lambda_i$'s  reduce to the constant coefficients in eqs.~(\ref{Hor_ops}), e.g. $\lambda_1=g_1m_0^2/(2\xi-\eta)$. 

Notice that the first two operators in the action~(\ref{EFTact_addendum})  have already been considered in ref.~\cite{Gleyzes:2013ooa}, while the remaining operators have been considered in refs.~\cite{Kase:2014cwa,Gao:2014soa}. However, in these latter works an explicit EFT action (in the form of the action~(\ref{EFTact_addendum})) has not been emphasized as well as an explicit mapping between these operators and a specific theory has not been worked out. In this respect our finding corresponds to new results.  
Finally, let us mention that although we wrote the operators in eqs.~(\ref{Hor_ops}) in terms of 3D quantities, following the 3+1 formalism employed in Ho\v rava gravity, one can always express them by means of 4D quantities by using the Gauss-Codazzi relation~\cite{Gourgoulhon:2007ue}.

It would be of interest to implement the contributions of these new operators in the equations for the perturbations evolved in EFTCAMB, in order to investigate their cosmological effects. We expect that their contribution becomes more important as the  cosmological scale becomes smaller. This is part of ongoing work~\cite{Frusciante:2016xoj}.

\section{Cosmological Parameters}\label{appCP}

In this appendix we report the $99.7\%$ C.L. constraints on the derived  cosmological parameters: $\Omega_b^0\,h^2$ the present day density parameter of baryons; $\Omega_c^0\,h^2$ the present day value of the cold dark matter density parameter;  $100\theta_{MC}$ which measures  the sound horizon at last scattering; $\tau$ is the Thomson scattering optical depth due to reionization; $n_s$ the power law index of the scalar spectrum; $ \mbox{ln}(10^{10}A_s)$ the log power of the primordial curvature perturbation, $H_0 \, \mbox{(km/s/Mpc)}$ the present time value of the Hubble rate, and $\Omega_m^0$ the dark matter density parameter today. We found that the constraints on these parameters for H2 are the same as in $\Lambda$CDM as reported in the following table. 
The reason for this is that, in H2 case, the Ho\v rava gravity parameters are constrained to be  very close to the GR limit so that  the  cosmological parameters fall back to their $\Lambda$CDM values.

\begin{table*}[h!]
\centering
\begin{tabular}{|c|c|c|c|c|c|}
\hline
\multicolumn{1}{|c|}{  }&
\multicolumn{5}{c|}{Bounds on cosmological parameters for $\Lambda$CDM and H2 cases}\\
\hline
\hline
Parameters & PLC & PLC+BG & PLC+CMBL & PLC+WiggleZ & all combined \\
\hline
\hline
$\Omega_b^0\,h^2$          & $0.02202\pm 0.008$		 	& $0.02216\pm 0.0007$ 			& $0.02210\pm 0.0008$ 			& $0.02182\pm 0.0008$  			& $0.02213_{-0.0007}^{+0.0008}$ \\[1mm]
$\Omega_c^0\,h^2$          	& $0.120\pm 0.008$		   		& $0.118\pm 0.004$    			& $0.119_{-0.006}^{+0.007}$    	& $0.123\pm 0.008$		   		& $0.119_{-0.003}^{+0.004}$ \\[1mm]
$ 100\theta_{MC}$        	& $1.041\pm 0.002$   			& $1.041\pm 0.002$			    	& $1.041\pm 0.002$			    	& $1.041\pm 0.002$		   		& $1.041\pm 0.002$ \\[1mm]
$\tau$                   		& $0.089_{-0.036}^{+0.047}$      	& $0.092_{-0.034}^{+0.040}$       	& $0.089_{-0.036}^{+0.041}$       	& $0.085_{-0.033}^{+0.037}$           	& $0.090_{-0.032}^{+0.037}$ \\[1mm]
$n_s$                    		& $0.959_{-0.020}^{+0.021}$      	& $0.963\pm 0.015$ 	      		& $0.961_{-0.021}^{+0.022}$       	& $0.953\pm 0.021$  	 		& $0.963\pm 0.015$ \\[1mm]
${\rm{ln}}(10^{10} A_s)$ 	& $3.088_{-0.069}^{+0.085}$      	& $3.090_{-0.069}^{+0.078}$       	& $3.085_{-0.064}^{+0.073}$       	& $3.087_{-0.063}^{+0.072}$ 		& $3.087_{-0.065}^{+0.071}$ \\[1mm]
$H_0$                    		& $67.2_{-3.4}^{+3.5}$              		& $68.0_{-1.6}^{+1.9}$               		& $67.7_{-2.9}^{+3.1}$               		& $65.8_{-3.1}^{+3.3}$ 			& $67.7_{-1.6}^{+1.8}$ \\[1mm]
$\Omega_m^0$               	& $0.316_{-0.044}^{+0.052}$     	 & $0.305_{-0.024}^{+0.022}$       	& $0.309_{-0.039}^{+0.042}$       	& $0.337_{-0.046}^{+0.051}$		& $0.309_{-0.021}^{+0.022}$ \\
\hline
\end{tabular}
\caption{Mean values and $99.7\%$ C.L. bounds on several cosmological parameters. Notice that these bounds do not change for the $\Lambda$CDM and H2 cases.}
\label{Tab:CosmologicalConstraintsH2}
\end{table*}

\begin{table*}[h!]
\centering
\begin{tabular}{|c|c|c|c|c|c|}
\hline
\multicolumn{1}{|c|}{  }&
\multicolumn{5}{c|}{Bounds on cosmological parameters for H3 case}\\
\hline
\hline
Parameters & PLC & PLC+BG & PLC+CMBL & PLC+WiggleZ & all combined \\
\hline
\hline
$\Omega_b^0\,h^2$          & $0.021^{+0.001}_{-0.001}$& $0.0218^{+0.0009}_{-0.0010}$ & $0.0218^{+0.0010}_{-0.0009}$ & $0.021\pm 0.001$  		& $0.0218_{-0.0009}^{+0.0008}$ \\[1mm]
$\Omega_c^0\,h^2$          	& $0.12^{+0.01}_{-0.01}$		& $0.119^{+0.005}_{-0.004}$    	& $0.118\pm 0.007$			    	& $0.123^{+0.008}_{-0.010}$		& $0.119_{-0.003}^{+0.004}$ \\[1mm]
$ 100\theta_{MC}$        	& $1.043^{+0.007}_{-0.004}$   		& $1.043^{+0.005}_{-0.003}$		& $1.043^{+0.006}_{-0.004}$		& $1.043^{+0.007}_{-0.004}$		& $1.044^{+0.004}_{-0.003}$ \\[1mm]
$\tau$                   		& $0.08_{-0.04}^{+0.05}$      	& $0.08 \pm 0.04$       	& $0.08 \pm 0.04$       	& $0.08_{-0.03}^{+0.04}$           	& $0.08_{-0.03}^{+0.04}$ \\[1mm]
$n_s$                    		& $0.96\pm 0.03$      	& $0.96\pm 0.02$ 	        	& $0.96\pm 0.02$       	& $0.96\pm 0.03$  	 	& $0.96\pm 0.02$ \\[1mm]
${\rm{ln}}(10^{10} A_s)$ 	& $3.09_{-0.08}^{+0.09}$      	& $3.09_{-0.08}^{+0.09}$       	& $3.08_{-0.07}^{+0.08}$       	& $3.10\pm 0.08$ 		& $3.07_{-0.06}^{+0.08}$ \\[1mm]
$H_0$                    		& $67\pm 5$	              		& $67\pm 2$               		& $68_{-3}^{+4}$               		& $65_{-4}^{+5}$ 			& $67_{-2}^{+1}$ \\[1mm]
$\Omega_m^0$               	& $0.31_{-0.05}^{+0.07}$     	 & $0.31\pm 0.02$       	& $0.30_{-0.04}^{+0.05}$       	& $0.33\pm 0.06$		& $0.30\pm 0.02$ \\
\hline
\end{tabular}
\caption{Mean values and $99.7\%$ C.L. bounds on several cosmological parameters in H3 case.}
\label{Tab:CosmologicalConstraintsH3}
\end{table*}


\end{document}